# A Partial Near-infrared Guide Star Catalog for Thirty Meter Telescope Operations

Sarang Shah[1], Smitha Subramanian[2], Avinash C. K.[2,3], David R. Andersen[4], Warren Skidmore[4], G. C. Anupama[2], Francisco Delgado[4], Kim Gillies[4], Maheshwar Gopinathan[2], A. N. Ramaprakash[5], B. E. Reddy[2], T. Sivarani[2], and Annapurni Subramaniam[2]

[1] India TMT Coordination Center, Bengaluru 56034, India; sshah1502@gmail.com
[2] Indian Institute of Astrophysics, Bengaluru 56034, India
[3] National Institute of Astrophysics, Optics, and Electronics, 72840, Mexico
[4] Thirty Meter Telescope International Observatory, 100 W. Walnut St., Suite 300, Pasadena, CA 91124, USA
[5] Inter-University Center for Astronomy and Astrophysics, Pune 411007, India

## Abstract

At first light, the Thirty Meter Telescope (TMT) near-infrared (NIR) instruments will be fed by a multiconjugate adaptive optics instrument known as the Narrow Field Infrared Adaptive Optics System (NFIRAOS). NFIRAOS will use six laser guide stars to sense atmospheric turbulence in a volume corresponding to a field of view of 2′, but natural guide stars (NGSs) will be required to sense tip/tilt and focus. To achieve high sky coverage (50% at the north Galactic pole), the NFIRAOS client instruments use NIR on-instrument wave front sensors that take advantage of the sharpening of the stars by NFIRAOS. A catalog of guide stars with NIR magnitudes as faint as 22 mag in the $J$ band (Vega system), covering the TMT-observable sky, will be a critical resource for the efficient operation of NFIRAOS, and no such catalog currently exists. Hence, it is essential to develop such a catalog by computing the expected NIR magnitudes of stellar sources identified in deep optical sky surveys using their optical magnitudes. This paper discusses the generation of a partial NIR Guide Star Catalog (IRGSC), similar to the final IRGSC for TMT operations. The partial catalog is generated by applying stellar atmospheric models to the optical data of stellar sources from the Panoramic Survey Telescope and Rapid Response System (Pan-STARRS) optical data and then computing their expected NIR magnitudes. We validated the computed NIR magnitudes of the sources in some fields by using the available NIR data for those fields. We identified the remaining challenges of this approach. We outlined the path for producing the final IRGSC using the Pan-STARRS data. We have named the Python code to generate the IRGSC as *irgsctool*, which generates a list of NGS for a field using optical data from the Pan-STARRS 3pi survey and also a list of NGSs having observed NIR data from the UKIRT Infrared Deep Sky Survey if they are available. *irgsctool* is available in the public domain on this GitHub public repository (https://github.com/sshah1502/irgsc), while the generated and validated IRGSC for the 20 test fields and additional Pan-STARRS Medium Deep Survey fields can be found on Zenodo.

*Unified Astronomy Thesaurus concepts:* Catalogs (205); Surveys (1671); Astronomy data analysis (1858); Observational astronomy (1145); Astronomical techniques (1684); Interstellar extinction (841); Star counts (1568)

## 1. Introduction

The Thirty Meter Telescope (TMT; Sanders 2013; Skidmore et al. 2015) will be one of the largest ground-based telescopes of the next decade, along with other large telescopes like the Giant Magellan Telescope (Bernstein et al. 2014) and European Southern Observatory's Extremely Large Telescope (Ramsay et al. 2014). Together, these so-called extremely large telescopes will all use adaptive optics systems (AOSs) for science observations (Hippler 2019), which will help to enhance our understanding of the cosmos. The TMT will have a multiconjugate adaptive optics (MCAO) system called the Narrow Field Infrared Adaptive Optics System (NFIRAOS;[6] Boyer & Ellerbroek 2016; Crane et al. 2018). The first-light science instruments on TMT—the Infrared Imager and Spectrograph (IRIS; Larkin et al. 2016) and the Multi-Objective Diffraction-limited High-resolution Infrared Spectrograph—will be fed by the AO-corrected light from NFIRAOS. The MCAO of NFIRAOS will be augmented by a laser guide star (LGS) facility, which will project up to nine lasers from the telescope into the sky to create artificial guide stars (Li et al. 2016). This facility and NFIRAOS's two deformable mirrors (DMs) and six LGS wave front sensors will enable high-quality and stable correction over a 2′-diameter field of view (FOV).

NFIRAOS on TMT will provide a real-time correction with a frequency of up to 800 Hz and will depend on both LGSs and natural guide stars (NGSs) for its operations. Because the laser beam is deflected by atmospheric turbulence both on the way up and on the way down, the position and motion of the laser beacon as it appears in the sky cannot be used to sense atmospheric tip/tilt. Furthermore, uncertainty in the exact height of the sodium layer makes the measurement of focus uncertain as well. Additionally, an MCAO system with multiple DMs and LGS wave front sensors can induce changes in plate scale (Flicker & Rigaut 2002). To correct for tip, tilt, focus, and the three plate-scale modes, the AO system needs feedback from three NGSs. Therefore, each NFIRAOS client instrument is designed to include three On-Instrument Wave front Sensors (OIWFSs) capable of sensing tip/tilt or tip/tilt/

---

[6] The main TMT mirrors will not be deformable, but NFIRAOS will correct the distortion before feeding the light to the science instruments.

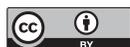

focus. These measurements are incorporated into the NFIR-AOS real-time controller, which commands the two DMs at a tip/tilt stage.

NFIRAOS and its client instruments are required to provide 50%سی sky coverage at the north Galactic pole. This means that NFIRAOS should meet the wave front error requirements for 50% of asterisms drawn from the density distribution of stars at the north Galactic pole. To meet this requirement, TMT determined that near-infrared (NIR) OIWFSs would be optimal; NFIRAOS would sharpen the NGS in the NIR, which will concentrate the light from the point sources onto a tiny area in the sky (roughly 12 mas in diameter) that will minimize the noise contribution from the sky background. The IRIS simulations using the Besançon Galaxy model (Robin et al. 2012) indicated that TMT would require NGS as faint as $J < 22$ to have a sufficient density of stars to meet the sky coverage requirements for NFIRAOS (Wang et al. 2012). Thus, a TMT-observable sky catalog of guide stars will thus be a critical resource for TMT operations. It will enable efficient planning and observing, fulfilling a role similar to that of the Guide Star Catalogs I and II (Lasker et al. 1990, 2008, 2008), which were created to allow for the acquisition and control of the Hubble Space Telescope (HST).

The TMT Infrared Guide Star Catalog (TMT-IRGSC) will be a star catalog consisting of point sources with NIR ($J$, $H$, and $K$) magnitudes as faint as 22 mag in the $J$ band in the Vega system, covering the entire TMT-observable sky. No catalog exists with objects as faint as $J = 22$ mag over the TMT as a whole observable sky. Hence, we compute the expected NIR magnitudes of stellar sources identified in various optical sky surveys using their optical magnitudes in multiple wave bands.

### 1.1. Previous Work on Developing TMT-IRGSC

Subramanian et al. (2013, hereafter S13) and Subramanian et al. (2016, hereafter S16) have already laid the foundation by developing a methodology to generate the IRGSC. However, they tested this methodology only on three test fields in the sky. They selected three test fields that had observed optical data from the MEGA-Prime camera of the Canada–France–Hawaii Telescope (CFHT; to compute the NIR magnitudes of the stars), which are available in three optical bands ($g$, $r$, and $i$), and the UKIRT Infrared Deep Sky Survey (UKIDSS) data (to compare the observed and computed magnitudes of the stars), which are available in $J$, $H$, and $K$ bands. S13 focused on laying the foundations to generate the guide star catalog by using blackbody models to compute the NIR magnitudes of the stars and separate the stars and the galaxies based on the spatial extent of the sources in the sky. These methods gave satisfactory results only in the $J$ band's 16–20 mag range, with a larger discrepancy in the computed $H$ and $K$ magnitudes that limited the number of sources for which the necessary faintness level was reached to less than the actual number required. S16 overcame this issue by using Kurucz (Kurucz 1992a, 1992b, 1993; Castelli & Kurucz 2003) and PHOENIX (Hauschildt et al. 1999a, 1999b) stellar atmospheric models (SAMs) and improving the methodology to compute the NIR magnitudes of the same stellar sources. The analysis described in S16 appeared to reach the required faintness levels for the TMT-IRGSC in the test fields.

Although S16 found that after using the SAMs the number of sources for which the NIR magnitudes were computed was significantly improved, nearly 30% of sources were brighter by 0.2–0.5 mag than observed, and the majority of the over-estimates were for stars with $T_{\rm eff} < 4000$ K. Therefore, they developed an optimal methodology to compute the sources' NIR magnitudes, divided into three stages. In the first stage, the interpolated Kurucz models having the parameters 3500 K < $T_{\rm eff}$ < 4000 K and log(g) < 3 and other interpolated Kurucz models with $T_{\rm eff}$ > 4000 K were applied to the stellar sources. The NIR magnitudes were then computed for the sources that had the difference in the reddened observed colors and the best-fitted model, which had less than twice the uncertainty in the observed colors. In the second stage, the interpolated PHOENIX models with 3000 K < $T_{\rm eff}$ < 3500 K and 3500 K < $T_{\rm eff}$ < 4000 K with log(g) < 3 were applied on the nonretrieved sources in the first stage. Finally, in the third stage, the two sets of sources were merged, and the method used to generate them was added as a flag in the catalog. This fine-tuned methodology was also applied to the publicly available Panoramic Survey Telescope and Rapid Response System (Pan-STARRS) reference data at that time.[7] The computed NIR magnitudes were validated using the Two Micron All Sky Survey (2MASS) NIR data. S16 found that the magnitudes were comparable to the observed 2MASS magnitudes, and the results were promising for the future production of the IRGSC from PS1 data, which covers most of the observable sky of the TMT and had additional longer optical bands, namely $z$ and $y$.

### 1.2. Overview

The main aim of this study is to apply the S16 methodology on the Pan-STARRS data in five optical bands, modify it to improve the results, and generate a partial IRGSC (for some selected fields) that will be similar to the final IRGSC for TMT.

The layout of the paper is as follows. Section 2 discusses the steps to generate and validate the generated IRGSC. The development of an optimal methodology to generate IRGSC using Pan-STARRS data is described in this section. The results obtained by applying the developed optimal method to 20 test fields are discussed in Section 3. The validation of the methodology by an alternate method is described in Section 4. In Section 5, we discuss the results obtained after applying the developed methodology to the Pan-STARRS data and how they satisfy the requirements of IRGSC. In Section 6, we provide the details of the Python package *irgsctool* (Shah & Subramanian 2024b), which can be used to generate the NIR magnitudes of any field in the PS1 survey using the Pan-STARRS optical data. We then summarize the work done in this study in Section 7 and present the plans for future development in Section 8. Appendices A and B discuss the nature of the generated and validated IRGSCs (Shah & Subramanian 2024a). We show additional figures related to developing our optimal methodology in Appendix C.

### 2. Steps to Generate and Validate the IRGSC

In this section, we describe the various steps involved in the computation of the NIR magnitudes of stellar sources from their optical magnitudes ($g$, $r$, $i$, $z$, and $y$ bands from the Pan-STARRS survey) and validation of the same. For this purpose, we selected 20 test fields across the sky and obtained the optical data from the Pan-STARRS 3pi survey (PS1). The 20 test fields we chose have a size of 30′ in radius or 0.785 deg² in area. These fields were selected to have data observed in five-band optical bands from the Pan-STARRS DR2 and NIR data from UKIDSS DR11, which were readily available. The observed NIR data from UKIDSS are used to compare the computed NIR magnitudes with their optical magnitudes, and based on the results, we modify and optimize the methodology for the generation of the IRGSC. The selected test fields are also located at various galactic latitudes throughout TMT's observable sky, which will help us understand the effect of interstellar extinction on the developed methodology (see Figure 1). In Table 1, we list the coordinates and the mean foreground reddening value along the line of sight of these fields, which is obtained using the reddening map of Schlegel et al. (1998) and later updated using the 2MASS data by Schlafly & Finkbeiner (2011), and we plot the positions of these fields in the sky in Figure 1. As we require point stellar sources in the generated catalog, we separate the stars and galaxies within the optical PS1 data set. The efficiency of the star–galaxy separation is verified using the Hubble Legacy Survey's archival data available for some test fields. To satisfy the astrometric information requirements, we obtained the Gaia data for these fields and cross-matched the stars to get their astrometric information, as the PS1 DR2 does not contain this information (however, the future releases may provide the astrometric information of the sources). We also generated a catalog of synthetic photometry and corresponding model parameters in the Pan-STARRS and UKIDSS filters using the Kurucz (Kurucz 1992a, 1992b, 1993; Castelli & Kurucz 2003) and PHOENIX (Hauschildt et al. 1999a, 1999b) SAMs. The developed optimal methodology is applied to the probable celestial sources in the test fields, and a partial IRGSC, with parameters similar to the final IRGSC, is produced. The steps in generating an IRGSC are shown as a flowchart in Figure 2, and each step is discussed in the subsections below.

---
[7] Since these data were not deep enough, as the input catalog went up to $i \sim 19$ mag only, S16 could not check the source density criteria.

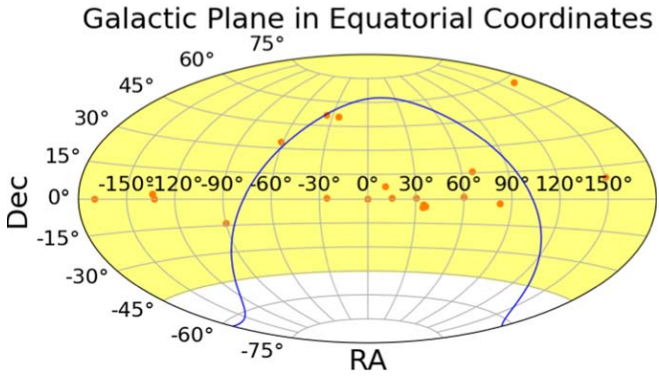

**Figure 1.** Location of the 20 test fields (as red circles) in the sky in Aitoff projection and equatorial coordinates. The TMT's sky coverage is the yellow shaded region, while the blue curves represent the Galactic plane ($b = 0°$).

### 2.1. Data

As shown in Figure 2, the first step is to obtain the data to generate and validate the IRGSC. This section describes the different data catalogs used in this study.

#### 2.1.1. Optical Data from Pan-STARRS for the Generation of IRGSC

Pan-STARRS is an innovative wide-field astronomical imaging survey (Chambers et al. 2016) that has its data processing facility developed at the University of Hawaii's Institute for Astronomy

**Table 1**
The Test Field Name and the Equatorial (Epoch J2000.0) and Galactic Coordinates of the 20 Test Fields Used in This Study

| Test Field | $\alpha$ | $\delta$ | $l$ | $b$ | $E(B-V)$ |
|---|---|---|---|---|---|
| TF1 | 227.26 | 0.0 | 359.27 | 47.24 | 0.04 |
| TF2 | 334.27 | 0.38 | 63.08 | −43.84 | 0.04 |
| TF3 | 60.00 | 1.25 | 188.72 | −36.53 | 0.26 |
| TF4 | 30.00 | 0.50 | 156.53 | −57.82 | 0.02 |
| TF5 | 11.16 | 7.83 | 120.00 | −55.00 | 0.04 |
| TF6 | 225.53 | 2.19 | 0.0 | 50.0 | 0.04 |
| TF7 | 269.93 | −13.48 | 15.00 | 5.00 | 0.98 |
| TF8 | 334.80 | 50.96 | 100.00 | −5.00 | 0.28 |
| TF9 | 324.09 | 51.47 | 95.00 | −0.50 | 2.48 |
| TF10 | 298.02 | 34.02 | 70.00 | 3.00 | 1.01 |
| TF11 | 0.00 | 0.00 | 96.33 | −60.18 | 0.02 |
| TF12 | 34.50 | −5.16 | 169.97 | −59.87 | 0.01 |
| TF13 | 36.25 | −4.50 | 171.65 | −58.22 | 0.02 |
| TF14 | 164.25 | 57.66 | 148.39 | 53.43 | 0.04 |
| TF15 | 66.75 | 15.86 | 180.08 | −22.32 | 0.58 |
| TF16 | 82.25 | −2.60 | 205.62 | −19.48 | 0.62 |
| TF17 | 189.83 | 0.00 | 296.33 | 62.71 | 0.01 |
| TF18 | 150.25 | 10.00 | 227.71 | 46.40 | 0.03 |
| TF19 | 15.00 | 0.90 | 127.47 | −61.89 | 0.02 |
| TF20 | 35.00 | −3.50 | 168.62 | −58.28 | 0.01 |

**Note.** The last column shows the median value of the foreground reddening toward the line of sight of these fields (Schlafly & Finkbeiner 2011).

(Magnier et al. 2020b; Waters et al. 2020) on the island of Maui, where the median seeing is 0″.83 (Chambers et al. 2016). The PS1 survey has several subsurveys for achieving different science goals. Out of these surveys, we use the data from the "3π Steradian Survey," to which the maximum survey time was dedicated and covering the sky north of $\delta = -30°$. These data are publicly available in five optical bands (*g, r, i, z,* and *y*) that contain the *mean* and *stack* photometry of the objects detected in its surveys in the AB magnitude system (Flewelling et al. 2020). Since the signal-to-noise ratio (SNR) and depth of the stack photometry, as well as completeness, are much higher than those of the mean photometry (Magnier et al. 2020a), we use the stack photometry (see Figure 3 for comparison between the mean and stack photometry). We obtained the stack photometry of the sources in the 20 test fields for a 30′-radius region of the sky centered on each test field. The sources that form a part of our input data set have at least one detection in all five optical bands. They have point-spread function (PSF) and Kron measurements in all five optical bands and have SNR > 5.

#### 2.1.2. NIR Data from the UKIDSS for the Validation of the Generated IRGSC

The UKIDSS was a set of five subsurveys with different photometric depths that scanned different parts of the sky depending on the scientific goals it had to achieve (Lawrence et al. 2007). UKIDSS used the Wide Field Camera (Casali et al. 2007) on the 3.8 m United Kingdom Infrared Telescope (UKIRT), which has an FOV of 0.21 deg$^2$, where each pixel of the camera subtends an angle of 0″.4 in the sky and the median seeing is 0″.6. We use UKIDSS-observed NIR data to validate the computed NIR magnitudes. The five subsurveys of UKIDSS—the Large Area Survey (LAS), Ultra-Deep Survey (UDS), Deep Extragalactic Survey (DES), Galactic Cluster Survey (GCS), and Galactic Plane Survey (GPS)—conduct observations in all of the five *ZYJHK* filters covering the wavelength range of 0.83–2.37 μm. Although UKIDSS has observed a sizable area and volume of the sky, the data cannot be used for IRGSC because they do not cover the entire observable sky of the TMT. However, as discussed in later sections, the observed NIR UKIDSS or VISTA data can be readily used for the fields close to the galactic plane, as the method to compute the NIR magnitudes from PS1 optical magnitudes fails to provide good results owing to the significant amount of extinction. We obtained the UKIDSS data for 30′-radius regions centered on the coordinates of our test fields to validate the computed NIR magnitudes.[8]

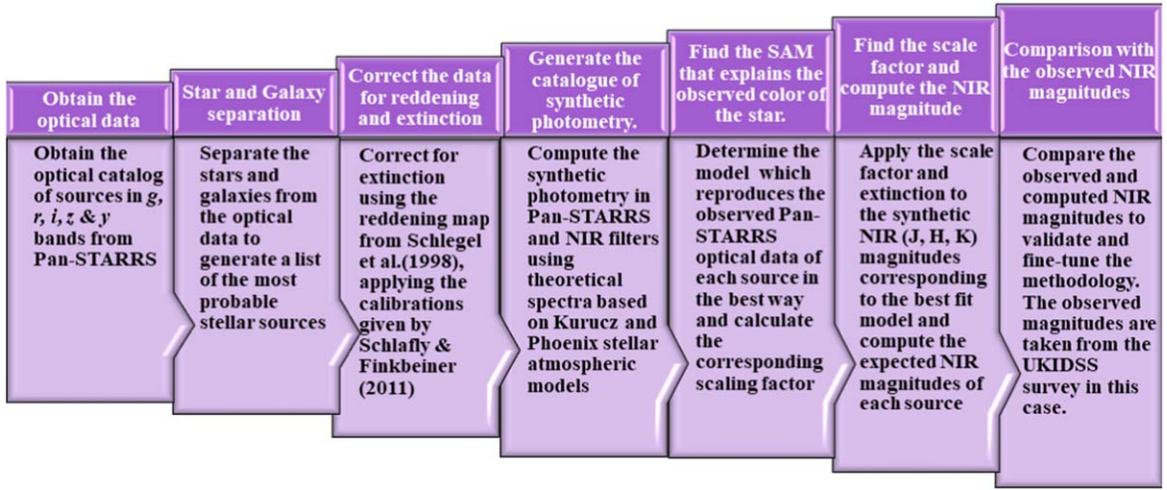

**Figure 2.** A flowchart representing generating an NIR guide star catalog.

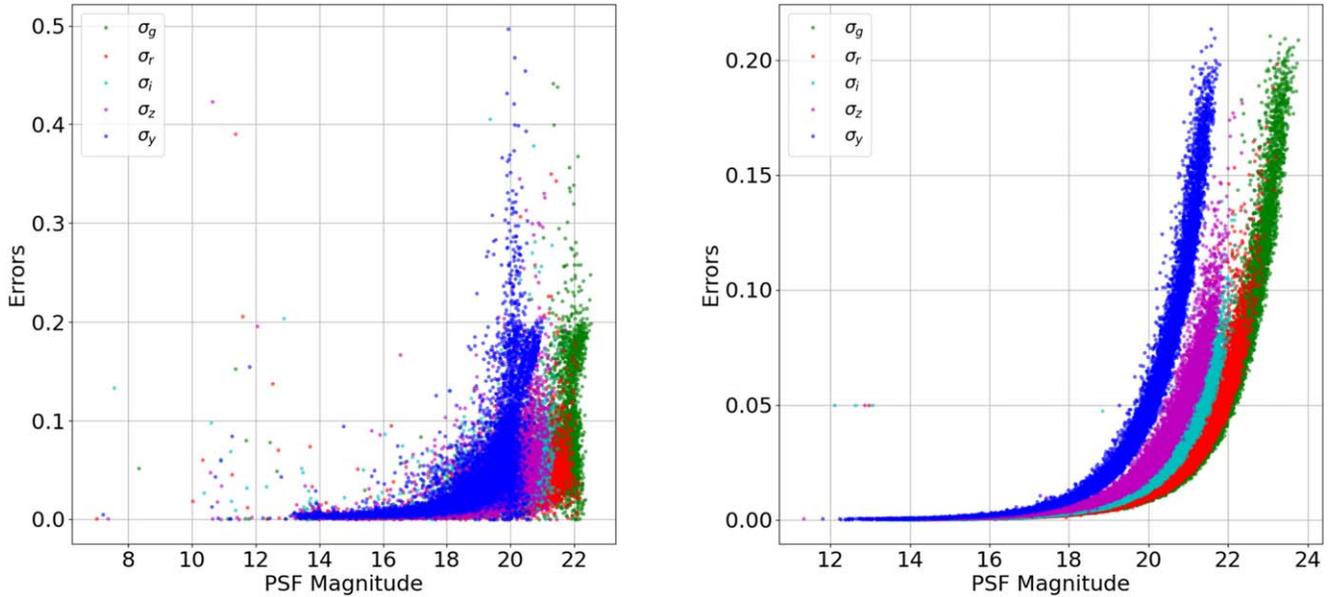

**Figure 3.** The left panel shows the error vs. magnitude plot in the mean photometric data, while the right panel shows the same for the stack photometric data. While the errors in each photometry increase exponentially in the fainter end, the maximum value of the error in the stack photometry is less than that in the mean photometry. Comparison of the *x*-axis also shows that the stack photometry is deeper than the mean photometry.

### 2.1.3. Astrometric Data from Gaia

Gaia is a space mission that has observed billions of stars in the Milky Way with its 1.4 m × 0.5 m space telescope since 2013 and provided their astrometric information. The primary aim of this mission is to generate a 3D model of the Milky Way by accurately measuring parallaxes, positions, and proper motions of all the stars. Gaia DR3 data were released in 2022 June and are based on the observations conducted between 2014 July and 2017 May (Gaia Collaboration et al. 2022). Among a variety of information, DR3 provides proper motions and parallaxes for billions of stars in the galaxy. In this work, we use the Gaia DR3 data for 30′-radius regions of the sky centered on each test field to provide the parallax and proper-motion information for the brighter sources in IRGSC. The Gaia data can be accessed via ESA's Gaia Data Archive or MAST CasJobs.

---
[8] We note that the UKIDSS magnitudes are petro magnitudes and not PSF magnitudes.

### 2.1.4. Hubble Source Catalog Data

The Hubble Source Catalog (HSC) is a catalog of sources that was released to help optimize science from the HST by combining several thousands of observations in the Hubble Legacy Archive (HLA) into a single master catalog (Whitmore et al. 2016). It includes WFPC2, ACS/WFC, WFC3/UVIS, and WFC3/IR photometric data generated using SExtractor software to produce the individual source lists. The space-based observations conducted using HST can achieve a higher spatial resolution than the ground-based telescopes, such that HSC can distinguish between the point and extended sources. Therefore, we use HSC to check the efficiency of the star–galaxy classification method that we employ on the PS1 data (see Sections 2.2 and 5.4 for more details). We use the catalog's latest version (v3), which has more sources over a wider area than the previous versions. After several rounds of post-processing of the images, the sources in the catalog have flags 0 for a point source and 1 for an extended source based on the photometry performed. Although we have tried to obtain the HSC data for the 30′-radius region of the test fields, these data are unavailable for all the test fields. The coverage is minimal for the test fields with HSC v3 data compared to the Pan-STARRS data.

### 2.2. Star–Galaxy Classification

The PS1 photometry we downloaded contains the PSF and the Kron photometry of the objects in all five optical bands. The PSF magnitude measures the total flux of a source fitted by a PSF model, which is a Gaussian. Thus, this model can estimate the brightness of the point sources accurately. Alternatively, the Kron magnitudes measure the total flux of an extended object inside the Kron radius ($R_K$). $R_K$ is defined as the first moment of the surface brightness light profile (Kron 1980). An $R_K$ of 2 or 2.5 contains a sufficient amount of the total flux of the galaxy to be a useful measure of total flux. This method can be used to determine the radius of faint extended sources. A Kron radius for an extended source will be larger than the width of the Gaussian of the PSF model fitted. Hence, the Kron and PSF magnitudes can be used to distinguish between the stars and the galaxies in the data. As we require only point stellar sources in the catalog, we separate the galaxies from the data set by using the relation

$$a_{PSF} - a_{Kron} < 0.05, \quad (1)$$

where "$a$" ∈ [$g$, $r$, $i$, $z$, and $y$] (Chambers et al. 2016).

### 2.3. Extinction Correction

Next, we correct the observed optical magnitudes of the probable stellar sources for the effects of interstellar extinction. We use the reddening map provided by Schlafly & Finkbeiner (2011) and the updated map based on Schlegel et al. (1998). The reddening values ($E(B-V)$) are obtained using the *dustmaps* Python package (Green 2018). The extinction coefficients in optical bands are calculated by using the relations given by Tonry et al. (2012), which are valid for $-1 < (g-i) < 4$. To validate the computed NIR magnitudes with the observed NIR magnitudes, we convert them to the apparent magnitudes by adding the extinction in the respective NIR filters. The NIR extinction values are obtained by multiplying the reddening obtained from *dustmaps* to the standard reddening-to-extinction $A/E(B-V)$ ratio of 0.709, 0.449, and 0.302 in $J$, $H$, and $K$ bands, respectively, assuming an extinction law based on Fitzpatrick (1999) and Indebetouw et al. (2005).

### 2.4. Synthetic Photometry in the Pan-STARRS and UKIDSS Filters

The parameters determining the nature of a star can be obtained by comparing the observed fluxes in the optical bands with the synthetic photometric fluxes for different types of stars predicted by the SAMs. The synthetic photometry of the stars in various filters can be computed from the synthetic spectra taken from different spectral libraries. To compute the synthetic magnitudes in the Pan-STARRS and UKIDSS filters, we use synthetic high-resolution spectra based on the Kurucz (Kurucz 1992a, 1992b, 1993), Castelli–Kurucz (Castelli & Kurucz 2003), and PHOENIX (Hauschildt et al. 1999a, 1999b) SAMs. SAMs provide expected high-resolution spectra from stars with different physical parameter combinations—($T_{eff}$, log ($g$), and [Fe/H]). As we have only photometric observations, we need to find the model magnitudes in our passbands ($g$, $r$, $i$, $z$, $y$, $J$, $H$, and $K$, in this case). To get the synthetic magnitudes, we calculate the effective stimulus (ES) by convolving the synthetic flux with the telescope response function (Pan-STARRS response functions are given in Tonry et al. 2012, and UKIDSS response functions are given in Hewett et al. 2006), multiplying it by the effective wavelength, integrating it across the wavelength range, and normalizing it (see Equation (2)). The flux in the SAMs is in units of $f_\lambda$ (erg s$^{-1}$ cm$^{-2}$ Å$^{-1}$). To compute the ES, we need to convert the flux to units of $f_\nu$ (erg s$^{-1}$ cm$^{-2}$ Hz$^{-1}$) (where $f_\nu = \frac{\lambda^2}{c} f_\lambda$). If the model flux of a star is $F_\lambda$ and the telescope's response function is $P_\lambda$, then the ES is given by

$$ES = \frac{\int F_\lambda P_\lambda \lambda d\lambda}{\int P_\lambda \lambda d\lambda}, \quad (2)$$

where $\lambda$ is the wavelength. Using Equation (2), we generate the synthetic photometry catalog in PS1 and UKIDSS filters and AB system of magnitudes (Oke & Gunn 1983), similar to the system used by PS1 (Tonry et al. 2012). In the synthetic photometry catalog, each set of magnitudes is associated with a [$T_{eff}$, log($g$), and [Fe/H]]. We use a Python package known as *pysynphot*,[9] which is an object-oriented synthetic photometry package from IRAF in Python (STScI Development Team 2013), to generate our synthetic photometry catalog. Table 2 shows the range of Kurucz, Castelli–Kurucz, and PHOENIX model parameters used in this study. Although the $T_{eff}$ goes up to 50,000 K for all the models, we have restricted the $T_{eff}$ to 10,000 K to increase the speed of the code to compute the NIR magnitudes. We also refine the grid axes of these model parameters by interpolating linearly between the parameters. These are called interpolated Kurucz, interpolated Castelli–Kurucz, and interpolated PHOENIX models, and we use these models in our work. Here we reduce the spacing between the temperature points to 62.5 K and between the log(g) and [Fe/H] points to 0.25 and 0.1 dex, respectively (see Table 2 for details on grid dimensions). While the Castelli–Kurucz models are the newer versions of the Kurucz models, the metallicity range they cover is less than that of the Kurucz models. We therefore combine the Kurucz model templates and the Castelli–Kurucz model templates.

---

[9] pysynphot.readthedocs.io

Table 2
Grid Size of the Various Stellar Atmospheric Models Used in This Study

| Stellar Atmospheric Model | $T_{\rm eff}$ (K) | log(g) (dex) | [Fe/H] (dex) |
|---|---|---|---|
| Kurucz | 3500–10,000 in steps of 250 K | 0.0–5.0 in steps of 0.5 dex | +1.0, +0.5, +0.3, +0.2, +0.1, 0.0, −0.1, −0.2, −0.3, −0.5, −1.0, −1.5, −2.0, −2.5, −3.0, −3.5, −4.0, −4.5, −5.0 |
| Interpolated Kurucz | 3500–10,000 in steps of 62.5 K | 0.0–5.0 in steps of 0.25 dex | 1.0 to −5.0 in steps of 0.1 dex |
| Castelli–Kurucz | 3000–10,000 in steps of 250 K | 0.0–5.0 in steps of 0.5 dex | 0.0, −0.5, −1.0, −1.5, −2.0, −2.5, +0.5, +0.2 |
| Interpolated Castelli–Kurucz | 3000–10,000 in steps of 62.5 K | 0.0–5.0 in steps of 0.25 dex | 0.5 to −2.0 in steps of 0.1 dex |
| PHOENIX | 2000–7000 in steps of 100 K, 7000–10,000 in steps of 200 K | 0.0–5.5 in steps of 0.5 dex | 0.0, −0.5, −1.0, −1.5, −2.0, −3.0, −3.5, −4.0, +0.3, +0.5 |
| Interpolated PHOENIX | 2000–10,000 in steps of 62.5 K | 0.0–5.0 in steps of 0.25 dex | 0.5 to −4.0 in steps of 0.1 dex |

### 2.5. Application of the Stellar Atmospheric Models to the Probable Stellar Sources

To estimate the parameters that give rise to the flux of a star in the data set, we fit the spectral energy distribution (SED) to the observed star. We do so by fitting the model colors to the dereddened observed colors and finding the scale factor to compute the NIR magnitudes. We verify the computed NIR magnitudes by comparing them with the observed NIR magnitudes from UKIDSS. In the following sections, we discuss how we generate an optimal method to generate an IRGSC. We used the data for the TF1 test field for this purpose, and later the best method was extended to all the other test fields.

#### 2.5.1. Using S16 Methodology and Finding Benefits of Using Five Optical Bands versus Three

Initially, we implement a formalism similar to S16, which was devised after checking for errors in the CFHT archival data. Since that method proved to be satisfactory in generating the NIR magnitudes, we apply it here to the PS1 data and fit the interpolated Kurucz models (we call them K0 models) to the observed color of the star. To do so, we first find the difference in the observed and model colors using the condition

$$|(\Gamma_{{\rm observed},p} - \Gamma_{{\rm model},p})| \leqslant 2 \times \sigma_{{\rm observed},p}, \quad (3)$$

where $\Gamma$ is the color for the filter combination "$p$" where $p \in n$ and "$n$" are all the possible dereddened color combinations using the five filters (*grizy*) of Pan-STARRS. Once a star satisfies this condition, we calculate the amount of deviation of each observed color from the model color and then calculate a quantity called $d_{\rm quad}$:

$$d_{\rm quad} = \sqrt{\sum_{p=1}^{p=n}(\Gamma_{{\rm obs},p} - \Gamma_{{\rm model},p})^2}. \quad (4)$$

The model with the lowest value of $d_{\rm quad}$ is the best-fitted model. To compute the expected NIR magnitudes from this best-fit model, we calculate the scaling factor (s.f.), which is defined as the difference between the observed and dereddened optical magnitudes and the model optical magnitudes:

$$\text{s.f.}_b = m_b - M_b, \quad (5)$$

where "$b$" denotes different optical bands. It is already shown in S13 and S16 that the s.f. is similar for all the bands. Nevertheless, we consider the *average value* of the s.f. in further analysis. We then add the model synthetic NIR magnitudes to the scale factor and account for extinction to get the apparent NIR magnitudes. The positions of the PS1 sources for which the NIR magnitudes are computed are positionally matched to the UKIDSS sources within 1″ to compare the computed and observed magnitudes. The value of 1″ was chosen because PS1 and UKIDSS telescopes have a spatial resolution of ∼1″.

The comparison plot of "observed NIR magnitudes" versus "difference in the observed and computed NIR magnitudes" was found to have a dual sequence by S16. The sources in the second sequence were found to have mean positive computed magnitudes brighter than the observed magnitudes and are the upper sequence of points in Figure 4. Based on their best-fit model parameters, they were suggested as cool giants by S16. Since the magnitude error of the sources in the input optical stack photometry is very small (typically ⩽0.2) as compared to the mean photometry, which has magnitude error in sources up to 0.5 (see, e.g., Figure 3), the majority of the sources do not satisfy Equation (3). Hence, we removed this condition and modeled all the sources with K0 models. Figures 4 and 5 show the comparison of the observed and computed NIR magnitudes color-coded according to the model parameters when all the sources are detected in *g*, *r*, and *i* bands and *g*, *r*, *i*, *z*, *y* bands, respectively. The sources not belonging to the primary sequence are cooler, metal-poor, and either compact or giant. When we include *z* and *y* bands in the modeling, although the displacement of the second sequence from the primary sequence decreases, it is still present. These stars are also cooler, metal-poor, and either compact or giant, but many metal-poor and cooler sources from Figure 4 find a better model. Thus, adding additional *z* and *y* bands helps model cooler sources better. Shown in Figure 6 is the SED of the best-fitted K0 model to a star when *g*, *r*, and *i* bands versus the *g*, *r*, *i*, *z*, and *y* bands are used to compute the NIR magnitudes. We can also see that adding *z* and *y* bands helps compute the NIR magnitudes. However, Figure 4 shows that K0 models are inefficient in the temperature range below 4000 K. Therefore, in addition to *z* and *y* bands, the sources must be modeled with PHOENIX models, as they are better in low-temperature regimes.

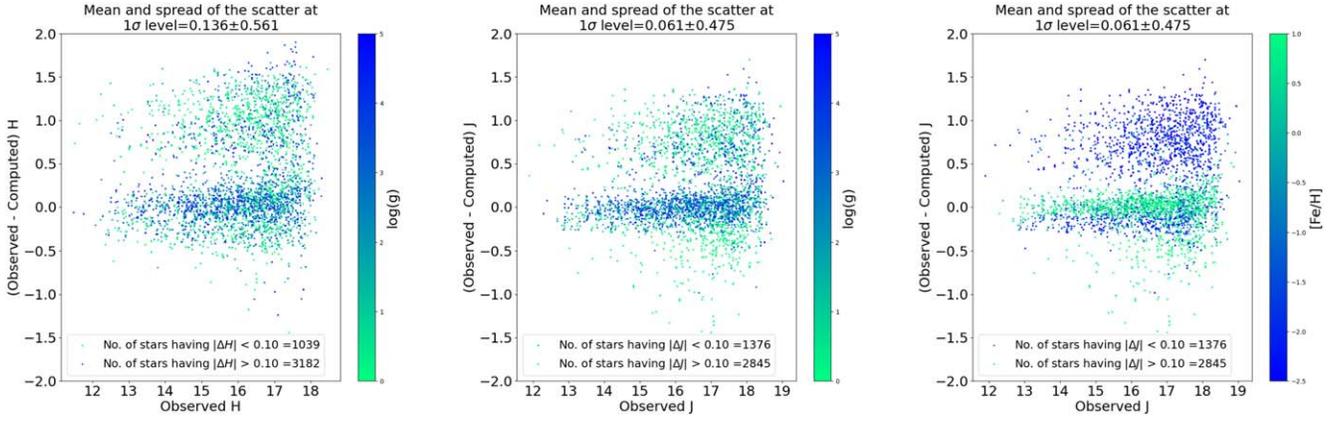

**Figure 4.** These plots compare the observed and computed *J* magnitudes of all the sources in the stack photometric data in *g*, *r*, and *i* bands. When we do not apply the condition in Equation (3), all the sources appear in the plot. We also find that the second sequence is prominent.

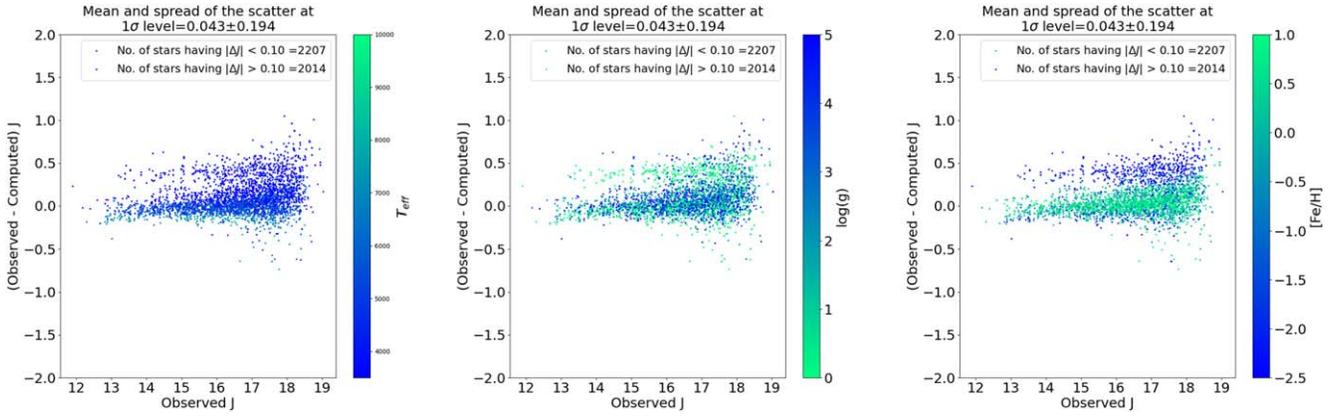

**Figure 5.** These plots compare the observed and computed *J* magnitudes of all the sources in the stack photometric data in *g*, *r*, *i*, *z*, and *y* bands. Although we find the presence of a second sequence, the density of the sources is far less than when only *g*, *r*, and *i* bands are used.

### 2.6. Finding an Optimal Model

Since the errors in the stack photometry are very small, we modify the strategy to find the best-fitting model. In particular, we start our search to find an optimal method by modeling all the sources using K0 models and calculating $d_{\rm dev}$ (see Equation (6)) for each source:

$$d_{\rm dev} = \sum_{p=1}^{p=n} (\Gamma_{{\rm obs},p} - \Gamma_{{\rm model},p})^2. \quad (6)$$

The NIR magnitudes are computed using the K0 model, which gives $d_{\rm dev,min}$. The comparison between the computed and observed NIR magnitudes is shown in Figure 7. In these plots, similar to the previous section, we find that K0 models are inefficient in low-temperature regimes. We also see the presence of dual sequences in the *H* and *K* bands. Hence, in the next step we model all the sources using Kurucz models with $T_{\rm eff} > 4000$ K (K1 models) only. The second row of Figure 7 compares the observed and computed NIR magnitudes when K1 models are used. Here the second sequence similar to the one in Figure 5 does not appear. However, we find that 1992 sources have $|(J_{\rm observed} - J_{\rm computed})| > 0.1$ and do not belong to the primary sequence. This is because not all the sources in the field are hot enough to be modeled efficiently with K1 models. In the *H* and *K* bands, this number is 2304 and 2688, respectively.

In Figure 8, we show the comparison of the observed and computed *J* magnitudes where the sources are color-coded according to the model parameters ($T_{\rm eff}$, log(*g*), [Fe/H]) and the cumulative deviation of the observed color from the best-fitted model color (dev). Similar plots for *H* and *K* are shown in Figure C1 in Appendix C. Most sources having absolute differences between observed and computed NIR magnitudes >0.1 are cooler, metal-poor, and compact. However, we also find that another population comprises cooler, metal-rich, and giant sources. The sources in this population also have a

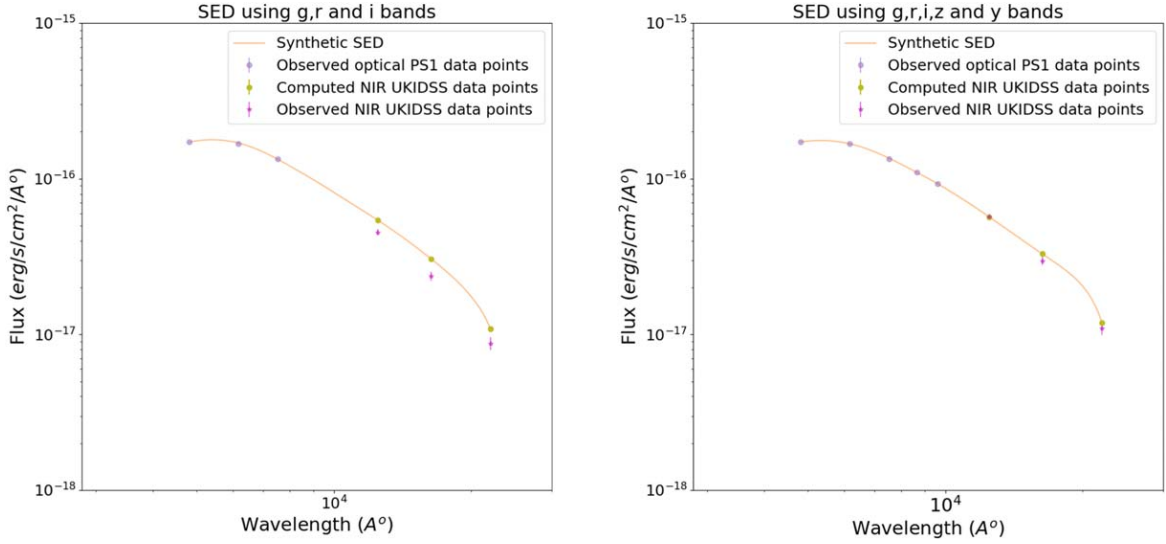

**Figure 6.** The left panel shows the SED (orange curve) of the best-fitted K0 model to a star when $g$, $r$, and, $i$ bands (shown as blue circles) are used. The parameters of the best-fitted model are (($T_{\text{eff}}$, $\log(g)$, [Fe/H]) = (5250.0, 5.0, 0.0)). In the right panel, the same star is fitted with a new K0 model when the $g$, $r$, $i$, $z$, and $y$ bands (also shown as blue circles). The best-fitted model parameters are (5125.0, 5.0, −0.8). In both panels, the magenta circles are the observed UKIDSS points in the $J$, $H$, and $K$ bands, while the yellow circles are the computed NIR magnitude points using the best-fitted model. The improvement in the accuracy of the computed NIR magnitudes, when additional $z$ and $y$ bands are used, is evident from this figure.

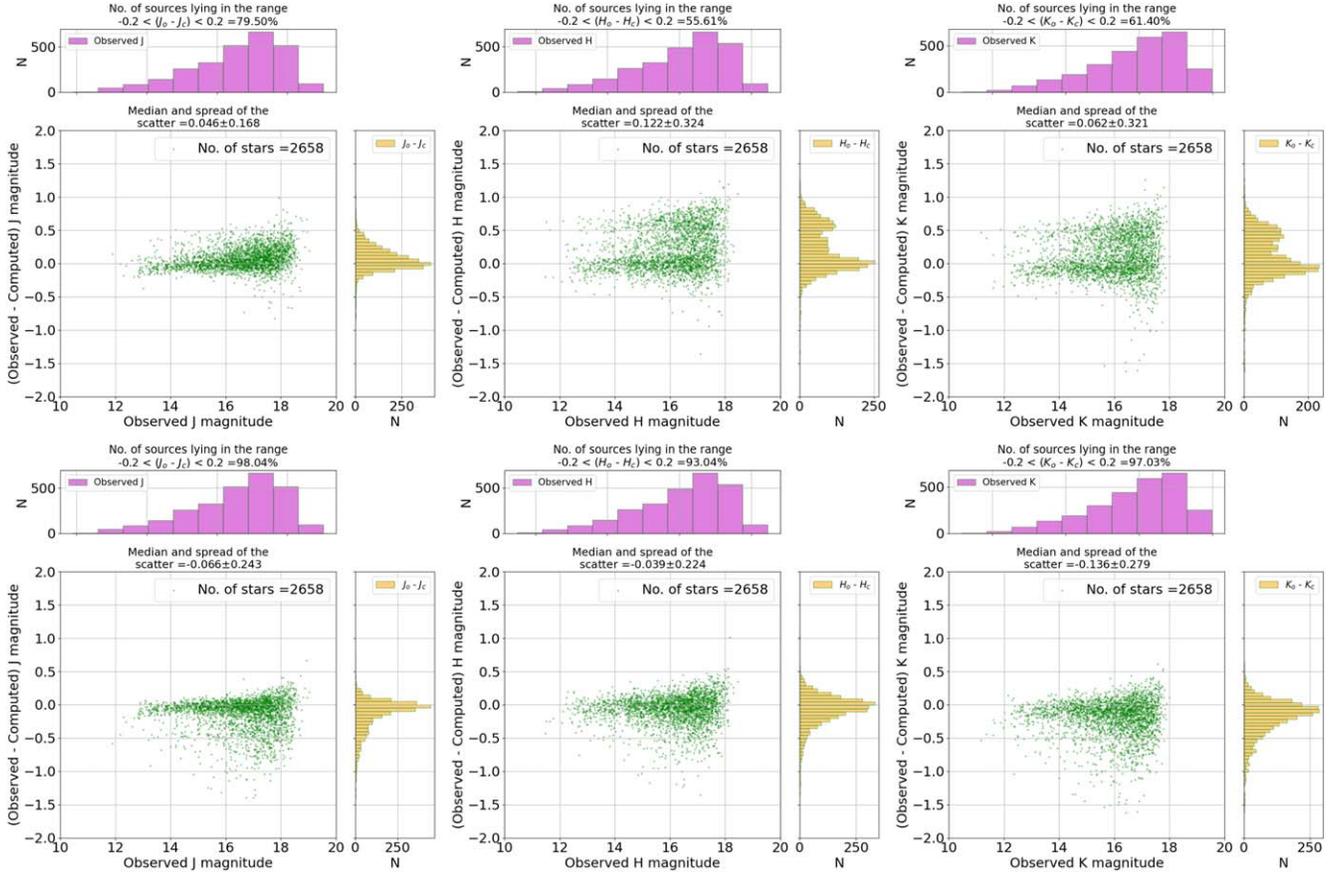

**Figure 7.** The plots compare the observed and computed $J$ magnitudes using the stack photometry data and when $d_{\text{dev,min}}$ is computed for each source. The panels in the top row are modeled using K0 models, while the panels in the bottom row are modeled using the K1 models.

smaller $d_{\text{dev}}$ value than the former population sources. Since PHOENIX models can model the cooler sources in a better way than the Kurucz models (as shown in the previous study by S16 and also by Bertone et al. 2004; Husser et al. 2013), we select sources having [$T_{\text{eff}} < 4500$ K, $\log(g) > 3.0$, [Fe/H] $< -1.5$, and dev $> 1.0$] and [$T_{\text{eff}} < 5000$ K, [Fe/H] $> -0.5$, $\log(g) < 3.0$, and dev $> 1.0$] and model them with PHOENIX model templates having $T_{\text{eff}} < 4000$ K (P0 models). In

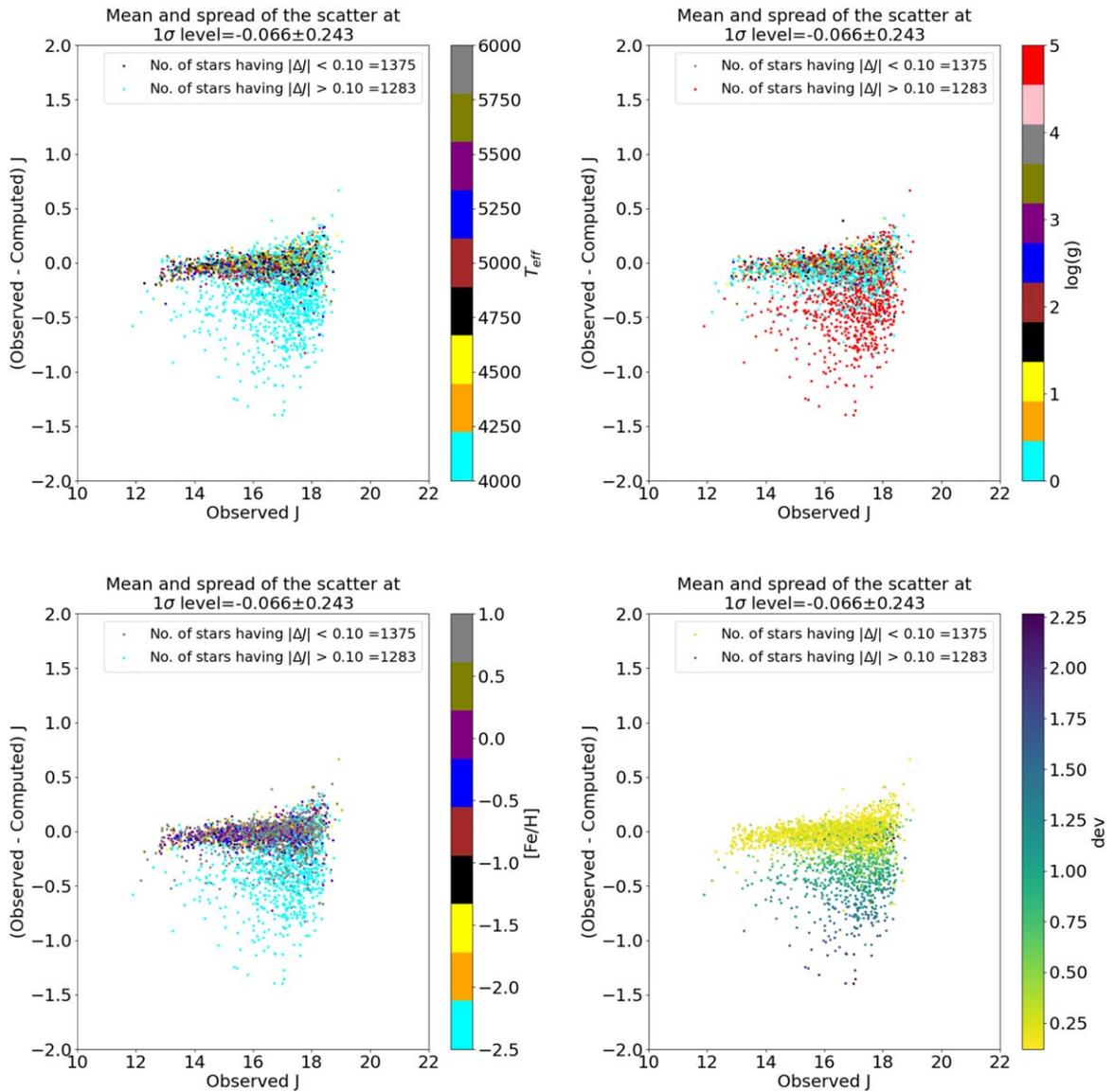

**Figure 8.** The panels show the comparison of the observed and computed $J$ magnitudes when all sources are modeled with K1 models and by calculating $d_{\rm dev,min}$ for each source (see Figure 7). Each panel shows the source color-coded according to the best-fitted model parameter and dev.

Figure 9, we show the comparison of the observed and computed $J$ magnitudes of these sources. Similarly, in Figure C2 in Appendix C, we show the comparison of the observed and computed $H$ and $K$ magnitudes. In these figures, we see the presence of multiple sequences in the plots, particularly in $H$ and $K$ bands, where the second and third sequences are highly displaced from the primary sequence. There are 658, 1034, and 966 such sources in the $J$, $H$, and $K$ bands, respectively, that do not belong to the primary sequence. This indicates that PHOENIX models with $T_{\rm eff} < 4000\,{\rm K}$ alone cannot compute the NIR magnitudes of certain ultracool (having $T_{\rm eff} < 3500\,{\rm K}$) sources effectively. Therefore, we model the *compacts* and *giants* using two specific groups of PHOENIX model templates.

From Figures 8 and C1, we know that the sources that do not belong to the primary sequence belong to two distinct populations of the sources: 1258 sources have [$T_{\rm eff} < 4500\,{\rm K}$, $\log(g) > 3.0$, [Fe/H] $< -1.5$], and 1994 sources have [$T_{\rm eff} < 5000\,{\rm K}$, [Fe/H] $> -0.5$, $\log(g) < 3.0$]. Generally, SAMs have issues in the low-temperature regime. Bertone et al. (2004) compared both Kurucz and PHOENIX SAMs in that regime and showed their degraded performance based on the theoretical fit of SEDs for a sample of 334 target stars along the whole spectral sequence. This was also found by Subramanian et al. (2016). Therefore, instead of selecting all PHOENIX models with $T_{\rm eff} < 4000\,{\rm K}$, we model the sources using selective PHOENIX models above 2800 K. In particular, the parameter range that we select is [$2800\,{\rm K} < T_{\rm eff} < 4000\,{\rm K}$, [Fe/H] $> -0.50$, $\log(g) < 3.0$] and [$2800\,{\rm K} < T_{\rm eff} < 5000\,{\rm K}$, [Fe/H] $< -1.50$, $\log(g) > 3.0$]. We call them the C1 and C2 models, respectively.[10] The sources largely displaced from the primary sequence now appear close to the primary sequence. The results are shown in Figure 10 for $J$ and Figure C3 in Appendix C for $H$ and $K$.

---
[10] We find that increasing the lower temperature limit of PHOENIX models from 2000 to 2800 K helps improve the accuracy of the computed NIR magnitudes by selecting better models and also helps in improving the computation speed.

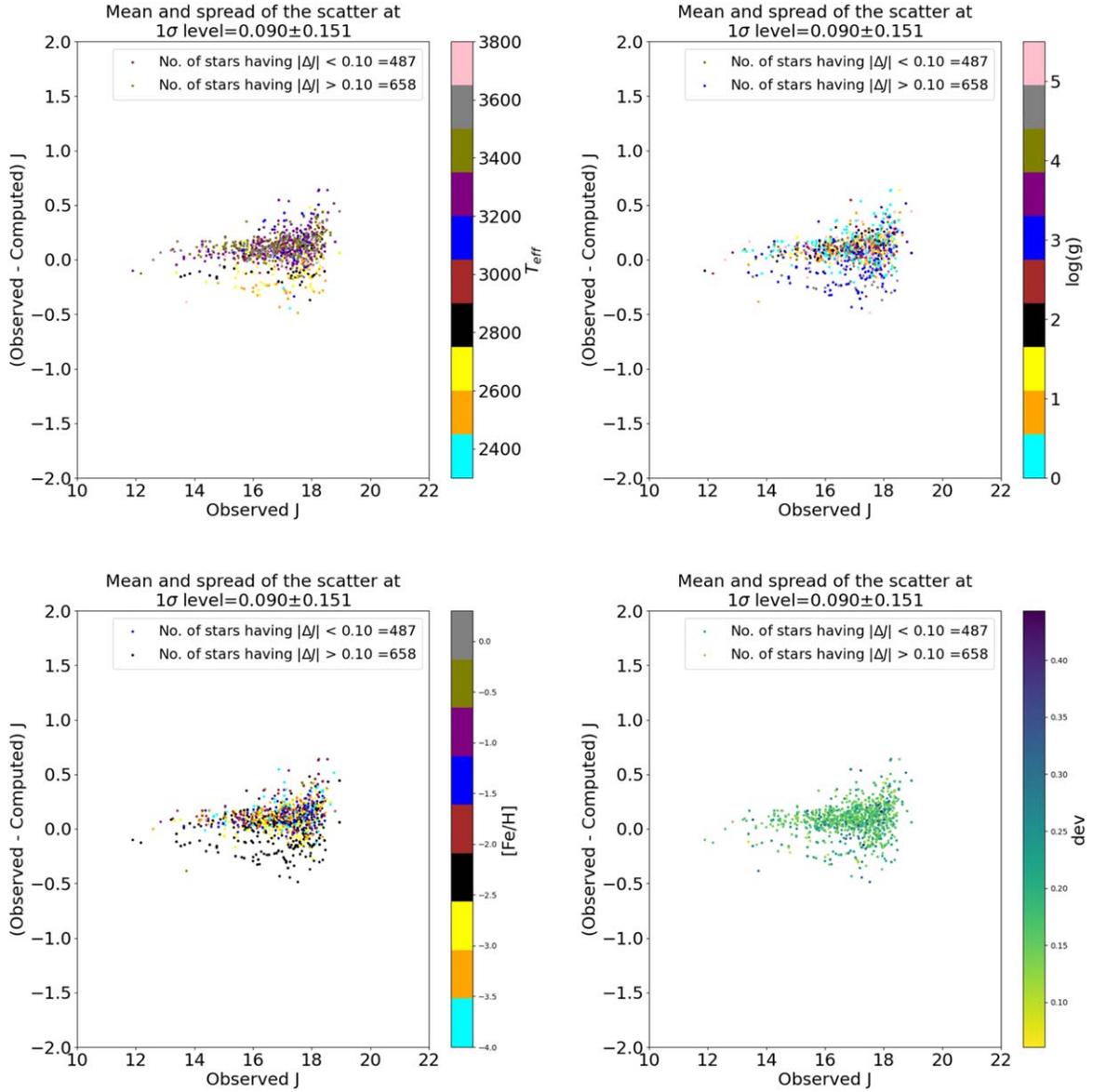

**Figure 9.** The panels show the comparison of the observed and computed *J* magnitudes when the scattered sources in the second row of Figure 7 are remodeled using P0 models.

*2.6.1. Combining Optimal Kurucz + PHOENIX Models*

In the previous section, we have seen that the method to compute the NIR magnitudes of the sources using the K0 models initially, and then K1 and then P0 models individually is not accurate. We find that remodeling of the sources that are scattered largely from the *J* band primary sequence in Figure 8 using C1 and C2 model templates not only improves the accuracy of the NIR magnitudes but also converts the random scatter into a sequence (see Figures 10 and C3). This sequence shifts toward the positive side, and the spread is larger in *H* and *K* bands. In the catalog, we flag these sources and suggest that they be used only if there are not enough sources to be used as guide stars in the field.

Thus, to suggest an optimal methodology for the generation of IRGSC, we model all the sources by combining K1, C1, and C2 model templates and computing $d_{\rm dev}$ for them. The results are shown in Figure 11. On comparing with the previous plots, we see that a proper selection of the range of $\log(g)$ and [Fe/H] values is crucial to reduce the spread of the scatter and reach the depth of $J = 22$ mag. Thus, this method can be called our *optimal* method to generate the IRGSC. In Figure 12, we show the properties of these sources where the median and the spread of the scatter in the *J* band are $0.03 \pm 0.13$. Similarly, in Figure C4 in Appendix C, we show the properties of the sources in *H* and *K* bands where the median and spread are $0.09 \pm 0.24$ and $0.03 \pm 0.26$, respectively. The median and spread in *J*, *H*, and *K* bands due to Kurucz models are $-0.02 \pm 0.10$, $-0.03 \pm 0.15$, and $-0.09 \pm 0.20$, respectively. Similarly, the median and spread due to the PHOENIX models for *J*, *H*, and *K* are $0.09 \pm 0.122$, $0.24 \pm 0.23$, and $0.15 \pm 0.25$, respectively. The outliers are, again, metal-poor, dwarf, and cooler sources. In addition, the width of the spread increases as the wavelength increases (i.e., increases from *J* to *K*). This effect can be seen across all the test fields (see the next section) and can be due to the limitation of the cooler SAMs. In the GitHub repository,[11] we provide the comparison scatter plots for the rest of the test fields after applying the optimal methodology on the stellar sources in these fields, and we present the result table in the next section.

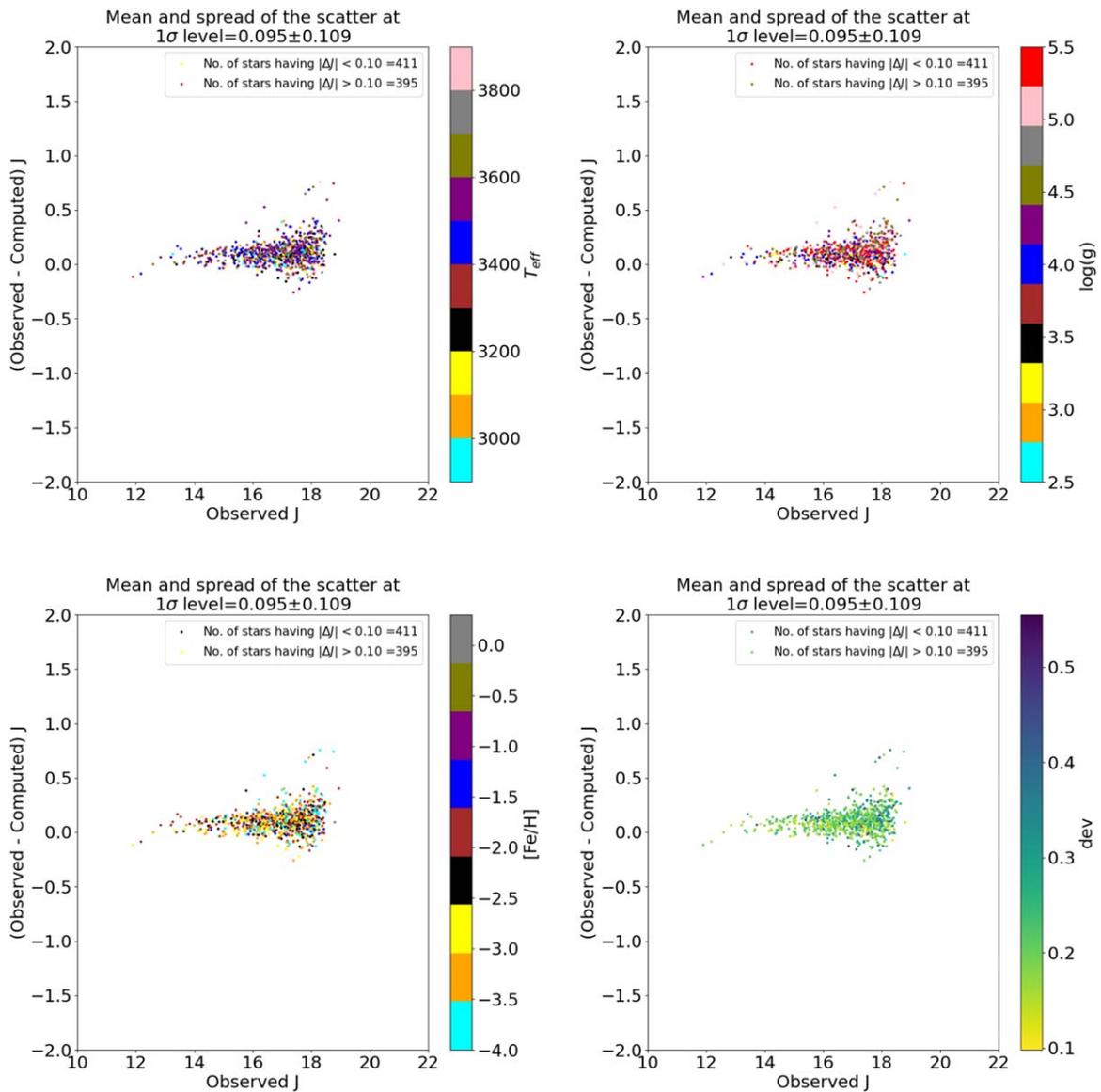

**Figure 10.** The panels show the comparison of the observed and computed *J* magnitudes when the scattered sources in the second row of Figure 7 are remodeled using C1 and C2 models.

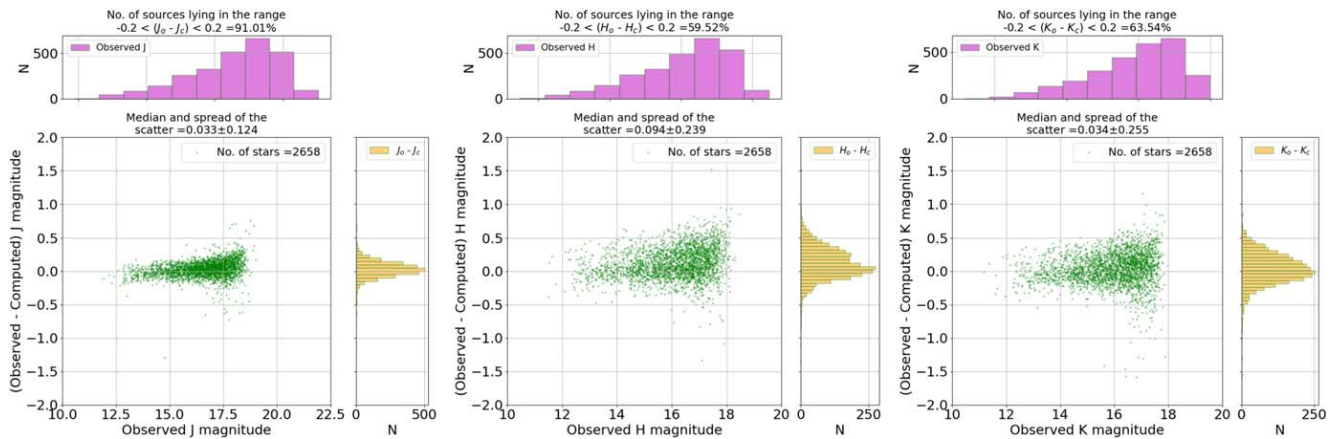

**Figure 11.** The panels compare the computed and the observed NIR magnitudes when the optimal method is applied on the stack photometry. There is a significant improvement in the accuracy of the computed NIR magnitudes (see the titles of each subpanel). The outliers are possible objects that are faint and cool.

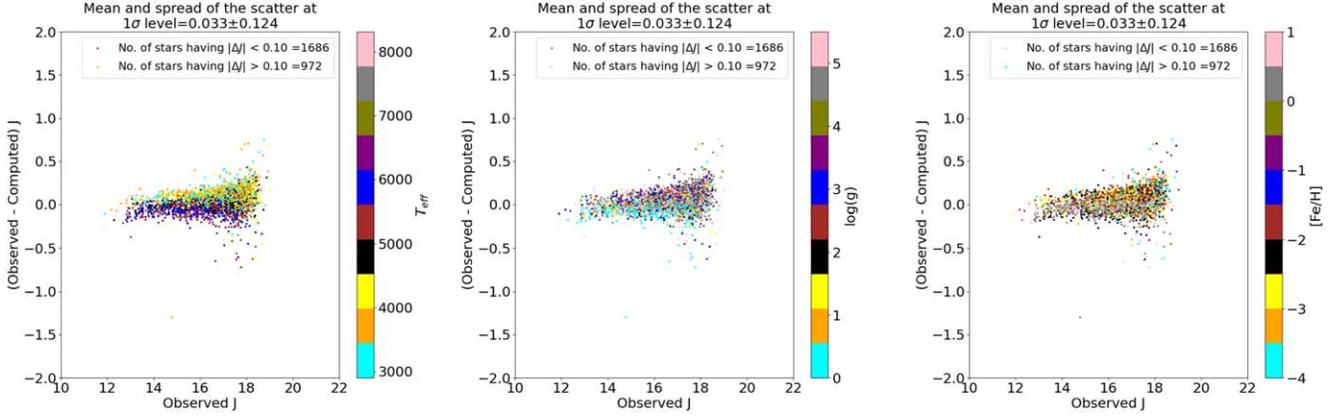

**Figure 12.** The panels show the properties of the sources in the plot, showing the comparison between observed and computed NIR magnitudes of the sources in the TF1 field when a combination of K1, C0, and C1 models is applied.

## 3. Results

We display the results obtained after applying the optimal set of Kurucz and PHOENIX models to the most probable stellar sources of the test fields. Table 3 shows the results for the test fields of size 30′ or 0.785 deg$^2$ region in the sky, where $N_{\rm sources}$ is the number of sources in the optical PS1 data for the particular field having at least two detections, $N_{\rm Cat}$ is the number of sources that are in our catalog,[12] $N_{\rm UKIDSS}$ is the number of sources in the UKIDSS NIR data to validate the computed NIR magnitudes, $i_{\rm comp}$ is the magnitude at which the $i$ band of PS1 data is 90% complete[13] (discussed in detail in the later part of the paper), $J_{\rm comp}$ is the magnitude at which the catalog is 90% complete, $J_{\rm faint}$ denotes the faintest magnitude of the source in the catalog, "Density" denotes the number of sources within the NFIRAOS FOV (as stated earlier, NFIRAOS requires three NGSs in its 2′-diameter FOV), and $\Delta_J$, $\Delta_H$, and $\Delta_K$ represent the median values of the distribution of the difference in the observed and computed $J$, $H$, and $K$, respectively, at the 1$\sigma$ level. We note that $N_{\rm UKIDSS}$ contains all the UKIDSS data sources with SNR > 5. This is a mixture of stars and galaxies, so this number does not equal the $N_{\rm Cat}$. The results and the IRGSC generated using the PS1 data and validated using the UKIDSS data are publically available on the GitHub repository.[14] The elements of TF10, TF13, TF14, and TF15 that are blank in the result table are due to the unavailability of the observed $J$- and $H$-band observed UKIDSS data for these fields. The asterisk represents the fields close to the galactic plane that suffer from high reddening and extinction. These fields are very crowded, so we obtained data for only 5′ radius compared to the 30′ radius for other test fields.

---

[11] https://github.com/sshah1502/irgsc
[12] It is to be noted that $N_{\rm Cat}$ is the same as the number of stars in the input catalog, as we separate the stars and galaxies. Consequently, the number of galaxies in the data will be $N_{\rm Cat}$—no. of stars.
[13] The 90% completeness of a catalog is the magnitude at which 90% of the sources are present.
[14] https://github.com/tmtsoftware/dms-irgsc/tree/main/generated_irgsc

## 4. Validation of the Results by an Alternate Method

We validate the computed NIR magnitudes using the *optimal color method* proposed above by an alternate method called the *flux method*. In this method, we treat $A_v$ and s.f. as free parameters and scale the model fluxes to the observed fluxes instead of colors (see Equation (7)). As per this method, the model with the least $\chi_r^2$ best fits a particular combination of s.f. and reddening. The advantage of this method is that it considers discrete extinction values for every star and its distance,

$$\chi_r^2 = \frac{1}{N - n_p}\sum_{i=1}^{N}\sum_{m=1}^{M}\left(\frac{f_{i,\rm obs} - 10^{-\rm s.f.}f_{i,m}10^{-0.4\left(\frac{A_\lambda}{A_v}\right)_i A_v}}{\sigma_{i,\rm obs}}\right)^2. \quad (7)$$

Here $N$ represents the total number of filters, $n_p$ denotes the number of free parameters (s.f. and $A_v$ in this case), $f_{i,\rm obs}$ is the observed flux, $f_{i,\rm model}$ is the model flux, $\sigma_{i,\rm obs}$ is the error in the observed flux, s.f. is the scale factor, and $(A_\lambda/A_v)_i$ is the ratio of the extinction coefficient of a particular filter to $A_v$. The best-fitted model is found by optimizing the values of the free parameters in the equation for a given $f_{\rm obs}$ and minimizing the $\chi_r^2$ hypersurface. A conventional approach involves using *nested for* loops to iterate through different values of these parameters, and the $\chi_r^2$ is computed by comparing each $f_{\rm model}$ for each $f_{\rm obs}$. However, the time complexity for this routine grows exponentially as the number of iterations increases. To address this issue, one common approach is to use *vectorization* of arrays. This method replaces iterative operations with vector operations, thereby reducing the time required to perform calculations and improving the speed of the fitting routine. Since the dimensions of the SAM array are huge, we implement this algorithm using optical photometry for the TF1 field on the *Google Colab*. Here we increase the computation speed by using the parallel computing ability of the GPUs enabled by the *PyTorch* Python package.

To vectorize the computation of the $\chi^2$, we construct a 5D array where the first axis (i.e., axis$_0$) represents the array of models fluxes, axis$_1$ represents the s.f. array (split into 200 bins in the range 20.0−25.0), axis$_2$ represents $A_v$ values (split into 200 bins in the range 0.0−3.0), axis$_3$ represents the observed fluxes, and axis$_4$ represents the filters. Since axis$_3$ has the observed fluxes of different stars, it is possible to find the best-

Table 3
The Results Obtained After Applying the Optimal Combination of K0, C1, and C2 Models to the Most Probable Stellar Sources in the Optical Data of the Test Fields

| Field | $N_{sources}$ | $N_{Cat}$ | $N_{UKIDSS}$ | $i_{comp}$ | $J_{comp}$ | $J_{faint}$ | Density | $\Delta_J$ | $\Delta_H$ | $\Delta_K$ |
|---|---|---|---|---|---|---|---|---|---|---|
| TF1 | 10376 | 4424 | 6441 | 20.94 | 19.64 | 22.65 | 4.91 | (90.82%) 0.030 ± 0.131 | (61.32%) 0.094 ± 0.254 | (65.69%) 0.034 ± 0.277 |
| TF2 | 13269 | 3945 | 4443 | 20.77 | 19.42 | 21.26 | 4.38 | (86.50%) 0.046 ± 0.175 | (64.29%) 0.116 ± 0.192 | (71.29%) 0.065 ± 0.195 |
| TF3 | 4628 | 2108 | 3298 | 20.36 | 18.64 | 21.00 | 2.34 | (90.65%) 0.051 ± 0.185 | (60.23%) 0.128 ± 0.214 | (70.37%) 0.037 ± 0.234 |
| TF4 | 6756 | 1778 | 2840 | 20.75 | 19.66 | 20.97 | 1.97 | (85.47%) 0.055 ± 0.162 | (53.51%) 0.152 ± 0.264 | (60.32%) 0.092 ± 0.303 |
| TF5 | 6447 | 1907 | 2763 | 20.67 | 19.31 | 22.10 | 2.11 | (88.86%) 0.071 ± 0.138 | (59.66%) 0.134 ± 0.245 | (67.05%) 0.069 ± 0.274 |
| TF6 | 8089 | 3327 | 4121 | 20.86 | 19.61 | 21.98 | 3.69 | (92.45%) 0.031 ± 0.140 | (62.51%) 0.085 ± 0.229 | (63.46%) 0.003 ± 0.261 |
| TF7⋆ | 2400 | 1700 | 4775 | 19.47 | 17.49 | 18.59 | 68.05 | (43.38%) −0.208 ± 0.268 | (58.87%) −0.068 ± 0.286 | (54.90%) 0.106 ± 0.327 |
| TF8⋆ | 1508 | 1281 | 1828 | 19.99 | 18.65 | 20.00 | 51.25 | (83.17%) −0.055 ± 0.178 | (66.35%) −0.052 ± 0.253 | (58.22%) −0.123 ± 0.276 |
| TF9⋆ | 1149 | 942 | 3103 | 19.82 | 17.99 | 19.33 | 37.69 | (13.74%) −0.436 ± 0.700 | (11.50%) −0.508 ± 0.877 | (17.78%) −0.371 ± 0.862 |
| TF10⋆ | 2370 | 1911 | 5921 | 19.68 | 18.01 | 19.07 | 76.45 | ⋯ | ⋯ | (42.92%) −0.210 ± 0.275 |
| TF11 | 4842 | 1492 | 4199 | 20.73 | 19.37 | 21.66 | 1.65 | (89.14%) 0.053 ± 0.151 | (56.26%) 0.128 ± 0.277 | (64.94%) 0.071 ± 0.295 |
| TF12 | 5388 | 1752 | 140204 | 20.82 | 19.81 | 22.57 | 1.94 | (89.01%) 0.060 ± 0.213 | (55.01%) 0.123 ± 0.456 | (61.05%) 0.061 ± 0.579 |
| TF13 | 5892 | 1762 | 41694 | 20.83 | 19.70 | 20.85 | 1.95 | ⋯ | ⋯ | (58.24%) 0.061 ± 0.370 |
| TF14 | 5828 | 1617 | 4211 | 20.73 | 19.35 | 21.89 | 1.80 | ⋯ | ⋯ | (52.95%) 0.089 ± 0.582 |
| TF15 | 6684 | 4036 | 12301 | 20.42 | 18.71 | 21.36 | 4.48 | ⋯ | ⋯ | (44.09%) 0.054 ± 0.373 |
| TF16 | 9560 | 6514 | 8259 | 20.39 | 18.94 | 21.37 | 7.23 | (70.81%) −0.132 ± 0.168 | (60.14%) −0.050 ± 0.259 | (59.28%) −0.061 ± 0.262 |
| TF17 | 8117 | 2873 | 6120 | 21.20 | 20.05 | 22.40 | 3.20 | (88.66%) 0.044 ± 0.148 | (56.98%) 0.119 ± 0.239 | (60.82%) 0.061 ± 0.257 |
| TF18 | 7243 | 2230 | 4057 | 20.76 | 19.57 | 21.85 | 2.47 | (91.92%) 0.040 ± 0.200 | (60.38%) 0.106 ± 0.287 | (67.06%) 0.063 ± 0.313 |
| TF19 | 6209 | 1675 | 1123 | 21.01 | 19.89 | 22.31 | 1.86 | (80.55%) 0.077 ± 0.237 | (50.60%) 0.174 ± 0.253 | (62.48%) 0.109 ± 0.220 |
| TF20 | 5510 | 1599 | 43196 | 20.66 | 19.58 | 22.22 | 1.77 | ⋯ | ⋯ | (61.46%) 0.057 ± 0.429 |

**Note.** The entries marked by the dash are for the test fields that do not have the observed UKIDSS data in the J and H bands, while a star indicates the fields close to the galactic plane. The % values in parentheses are the percentage of sources within 0.2 mag of the absolute difference in the observed and the computed NIR magnitudes, and the density is computed for the NFIRAOS FOV.

fit parameters for multiple stars in a single run. The indices of the minimum $\chi^2$ value across axis$_0$, axis$_1$, and axis$_2$ provide the best-fitted model, s.f., and $A_v$ information.

We apply the flux method to the Pan-STARRS optical data of the TF1 field in two ways. First, we keep $A_v$ as a free parameter within the abovementioned range. Since a wider range of $A_v$ may lead to overfitting and unrealistic prediction of $A_v$, next we choose a tighter range for $A_v$ where $A_v \in \{A_v', 3\sigma A_v'\}$ and $A_v'$ is taken from Schlafly & Finkbeiner (2011). We plot the results in Figure 13; the left panel shows the comparison of the difference in the computed and observed J magnitudes for the TF1 field when $A_v$ is kept free but bounded by a $3\sigma$ limit, whereas the right panel shows the same but when the $A_v$ is kept free within the range [0.0, 3.0]. It is to be noted that the mean value of $A_v$ for the TF1 field is 0.04 (refer Table 1). Similarly, Figure C5 in Appendix C shows this comparison for H and K bands, respectively. The color bar shows the $A_v$ values that each source takes to minimize the $\chi_r^2$. We find that the predicted NIR magnitudes are better computed when $A_v$ is fixed by $3\sigma$ compared to when $A_v$ is left unconstrained. Although placing every star at a different distance by treating s.f. and $A_v$ as free parameters is realistic, in the case of NIR bands the computed magnitudes when $A_v$ is bounded and unbounded by the $3\sigma$ limit do not differ significantly from the computed NIR magnitudes using the optimal color method where $A_v$ is fixed. This is the case especially for sources that show larger deviations from the observed NIR magnitudes. In addition, keeping $A_v$ unbounded gives a larger scatter concerning the observed magnitudes, meaning that $A_v$ takes any value to minimize the $\chi_r^2$. Since computing the NIR magnitudes by keeping $A_v$ and s.f. free is computationally more extensive and time-consuming, we suggest that the optimal color method can generate IRGSC.

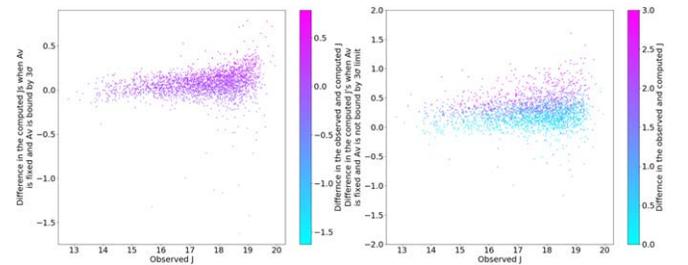

**Figure 13.** The left panel shows the comparison of the difference in the computed and observed J magnitudes for the TF1 field when $A_v$ is kept free but bounded by a $3\sigma$ limit, whereas the right panel shows the same but when the $A_v$ is kept free within the range [0.0, 3.0]. The mean value of $A_v$ for the TF1 field is 0.04. The color bar shows the $A_v$ values that each source takes to minimize the $\chi_r^2$.

## 5. Discussion

We have devised a method that utilizes observations of sources in five optical bands in the PS1 3pi survey DR2. This method is devised to reach the required faintness of $J = 22$ mag, achieving the source density criteria in the NFIRAOS FOV while maintaining the accuracy of the computed NIR magnitudes. In this section, we discuss the results in detail.

### 5.1. Accuracy of the Computed NIR Magnitudes and the Faintness Achieved

The results shown in Table 3 are obtained after applying the combination of K0, C1, and C2 models to the most probable stellar sources in the fields. We aimed to compute the NIR magnitudes for the probable stellar sources in each test field by maintaining the required accuracy of 0.2 mag and reaching up to $J = 22$ mag. We have observed that the faintest $J$ value computed for many test fields is approximately 22 mag. The median difference between observed and computed $J$ values for all test fields is $-0.43$ to $0.07$ mag, including fields close to the galactic plane. If we exclude the fields TF7, TF8, TF9, and TF10, i.e., the fields located close to the galactic plane, the range of difference is between 0.03 and 0.07 mag, which is acceptable. The main sources of error in computed NIR magnitudes are the input optical magnitudes and the error in reddening. The error in computed $J$ values is typically in the range of millimagnitudes to a few tenths of a magnitude. However, the spread is greater in the $H$ and $K$ bands, possibly due to poor modeling of ultracool, metal-poor dwarfs by the SAMs.

It is essential to meet the source density criteria for the NFIRAOS FOV and source faintness. The condition of the source density criterion is at least three stars in the 1′-radius NFIRAOS FOV. This number translates to 3440 stars deg$^{-2}$. In Table 3, excluding the fields located close to the galactic plane, the test fields TF1, TF2, TF6, TF12, TF15, TF16, and TF17 satisfy our requirement of the source density and the others do not. The achieved source density in the fields where the requirement is not met is generally quite close to the requirement but is fundamentally limited because these fields are located at a high galactic latitude and have less source density in the input PS1 data themselves (see $N_{\text{Cat}}$ column in Table 3 and Section 5.2).

### 5.2. 90% Completeness of the Catalog

The 90% completeness of a catalog is the magnitude at which 90% of the sources are present. This number is important because it gives us an idea of the magnitude depth of a catalog and the number of sources reaching that depth. When we plot a distribution of the sources, 90% completeness also resembles the bin in which a maximum number of sources lie. We compare the distribution of stars in the input PS1 catalog with the synthetically generated catalog of stars using the latest version of the Besançon Galaxy model of stellar population synthesis (Czekaj et al. 2014). This model can be accessed through a web query after creating an account on their official website.[15] Here, to create a new simulation, one has to set up the model by referring to the version of the model and the photometric system to be used. We used the SDSS + 2MASS photometric system as PS1, and UKIDSS photometric systems are unavailable on the web page and resemble the former systems. Since we are only interested in the star counts toward the line of sight of the field, for simplicity we do not select a model with the kinematics of stars included. In this way, we generate the simulated stellar counts toward the center of all the test fields without any constraint on the magnitude and color range. The results table lists the completeness value of the generated IRGSC for all our test fields.

In Figure 14, we plot the distributions of $i_{\text{PSF}}$ at various stages for the TF1 field Pan-STARRS data. The top left panel shows the histogram of the $i$ band when the data set contains the sources detected only in the $i$ band. The top right panel shows the distribution of the $i$ band when the sources detected in all five bands are considered. In the bottom left panel we show the $i$-band distribution of the probable stellar sources obtained after the star–galaxy classification applied only in $i$ band. The 90% completeness of the distribution is at $i \sim 21.27$. The bottom middle panel shows the same but when the star–galaxy classification is applied to all five optical bands. This time, the 90% completeness reduces to $i \sim 20.94$. Thus, if we consider the data set that includes the sources detected only in the $i$ band, the number of sources and 90% completeness are more than for the sources detected in all five bands. Finally, in the bottom right panel we show the distribution of the $J$ magnitude of the sources in IRGSC computed using the optimal methodology developed in this work. The 90% histogram of this distribution is at $J \sim 19.64$. All the histograms in the bottom row are plotted on top of the histogram of the Besançon model that shows the predicted distribution for stellar sources for the TF1 field. Thus, we can see that our condition that a source must have at least one detection in each band affects the 90% completeness and source density of the input optical data, thereby affecting the 90% completeness of the generated IRGSC.

### 5.3. Applying the Optimal Methodology on Sources Detected in g, r, and i Bands Only

We have seen that the required depth of $J = 22$ mag and the required source density are not satisfied in some of the test fields, such as TF11, TF14, TF19, and TF20, due to the limitation of the Pan-STARRS data. Hence, we apply our optimal methodology for the sources in these test fields detected only in $g$, $r$, and $i$ bands and see the results. We note that the star–galaxy separation is based on these three bands only. Naturally, the number of sources in the IRGSC will increase, but at the cost of the accuracy of the computed NIR magnitudes. Overall, we find an increase of up to 35% in the sources, and the $J_{\text{comp}}$ goes beyond 21.0 ($J_{\text{comp}}$ is 19.64 when all five bands are used; see Table 3). Both of these changes are sufficient to satisfy our requirements, but the accuracy of the computed NIR magnitudes decreases. There is a reduction of up to 16% in the number of sources lying within $-0.2$ to $+0.2$ of the difference in the computed and observed $J$ mag. The same number is up to 10% for the $H$ band and 11% for the $K$ band, respectively.

---

[15] https://model.obs-besancon.fr

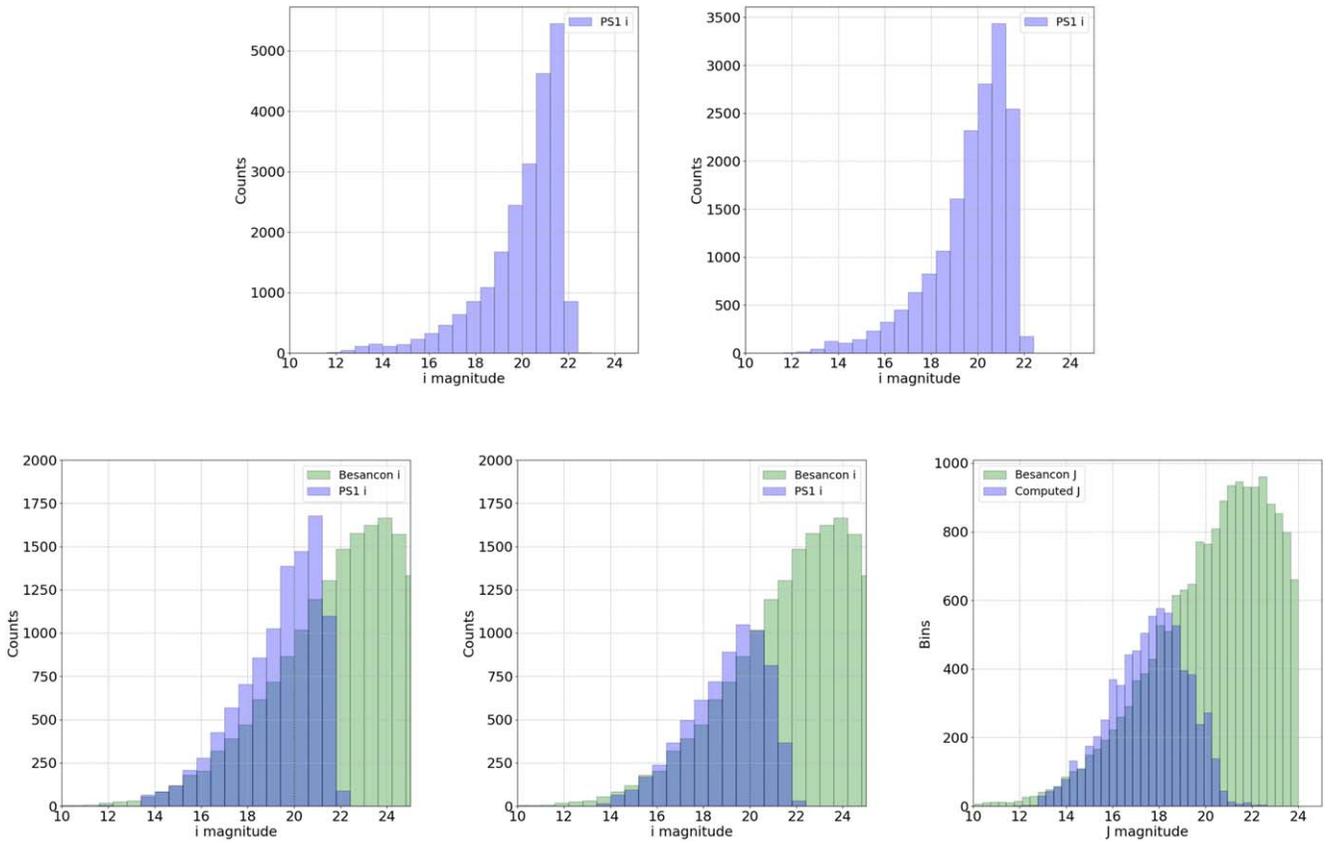

**Figure 14.** Top left: histogram of $i_{\rm PSF}$ when the input data set contains the sources that have detections only in the $i$ band. Top right: histogram of $i_{\rm PSF}$ when the data set contains sources having detections in all five PS1 bands. Bottom left: histogram of $i_{\rm PSF}$ after star–galaxy classification applied only in $i$ band on the sources detected in all five bands and that have SNR $\geqslant 5$. Bottom middle: histogram of $i_{\rm PSF}$ for the sources after applying the star–galaxy classification in all five bands and the sources that have SNR $\geqslant 5.0$. Bottom right: when the sources have SNR $> 5$, the distribution of the computed $J$ is detected in all five bands, and the star–galaxy classification is also applied in all five bands. It is modeled using the optimal methodology developed in the work. All the distributions are plotted using the Pan-STARRS data for the TF1 field. The distributions in the bottom row are plotted on top of the Besançon model distribution for the TF1 field (shown in green).

### 5.4. Testing the Efficiency of Star–Galaxy Separation

The result table is obtained after assuming that the modeled sources are stellar sources obtained after applying Equation (1). However, to test the efficiency of the star–galaxy classification method, we must compare the sources classified as stars and galaxies in the PS1 data with a catalog containing known stars and galaxies in a particular region of the sky. We chose the HST observational data. The observations by the HST are archived in the HSC (Whitmore et al. 2016) by combining the tens of thousands of visit-based source lists in the HLA into a single master catalog. We select the Extended Groth Strip (EGS; Davis et al. 2007) region in the sky ((R.A., decl.) = (214.8250, 52.8239)), which is an extension of an HST Groth Strip survey that was carried out in 1994 by the WFPC team (Groth et al. 1994; Rhodes et al. 2000). This region has low extinction and low Galactic infrared emission. Thus, it has deep observations across various wavelengths. A portion of the EGS (70.′5 × 10.′1 in area) was also imaged by the HST from 2004 June to 2005 March. We obtain these data from the latest version of the HUBBLE Source Catalog v3.1 (Whitmore et al. 2016) and separate the stars and galaxies using the *flag* value for stars and galaxies.[16] Similarly, we obtain the PS1 data for the same region from the Pan-STARRS 3pi survey such that it contains sources detected in all five optical filters having SNR $> 5$. Before checking the efficiency of the star–galaxy classification, we cross-match the sources positionally to within 1″. We found several sources with more than one HSC counterpart within 1″ search distance. We remove these sources from our analysis because it would mean that the PS1 magnitudes are a combination of the individual magnitudes of these sources.[17] We then classify the sources as stars and galaxies using Equation (1) and check for the flags of these sources in the HSC data set. The results are displayed in Table 4, where the first column represents the field name and the second ($N_{\rm HSC}$) and third ($N_{\rm PS1}$) columns represent the number of sources in HSC and PS1, respectively. The fourth column ($N_m$) represents the number of sources in PS1 that find more than one counterpart in the HSC within 1″, and the fifth column ($N_c$) represents the number of common sources in both the catalogs after removing $N_m$. Then, the next four columns represent the number of stars and galaxies in the HSC and PS1 data, where the subscript "gx" stands for galaxies and the subscript "st" stands for stars. Finally, in the last four columns we show the number of PS1 stars that are HSC galaxies

---
[16] Flag $= 0$ for stars and any other flag for saturated and nonstellar sources.
[17] These sources are also removed when we apply our star–galaxy classification criteria on the five optical bands.

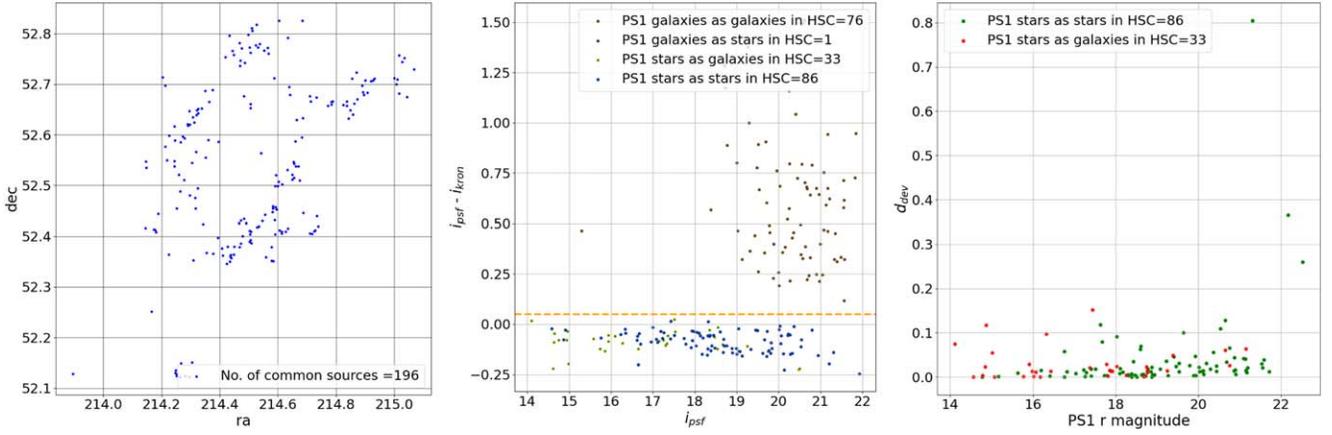

**Figure 15.** In the left panel we depict the equatorial positions (epoch J2000.0) of the sources common to PS1 and HSC within the EGS field. The middle panel illustrates the relationship between PS1's $i_{PSF}$ and the difference between $i_{PSF}$ and $i_{Kron}$. The orange horizontal line at 0.05 indicates the classification threshold of distinguishing between stars and galaxies in the PS1 data set (as defined in Equation (1)). Each data point is color-coded to denote its classification consistency: green signifies galaxies classified as such in both PS1 and HSC, orange indicates sources identified as galaxies in PS1 but as stars in HSC, olive represents stars in PS1 but galaxies in HSC, and blue denotes sources classified as stars in both PS1 and HSC. Lastly, the right panel showcases the metric $d_{dev}$ for sources classified as stellar in PS1 but categorized as either stars (green points) or galaxies (red points) in HSC.

**Table 4**
The Results After Comparing the Stars and Galaxies in the PS1 Data with the Common Sources in the HSC Data

| Field | $N_{HSC}$ | $N_{PS1}$ | $N_m$ | $N_c$ | $N_{st,HSC}$ | $N_{gx,HSC}$ | $N_{st,PS1}$ | $N_{gx,PS1}$ | PS1$_{st}$/HSC$_{gx}$ | PS1$_{gx}$/HSC$_{gx}$ | PS1$_{st}$/HSC$_{st}$ | PS1$_{gx}$/HSC$_{st}$ |
|---|---|---|---|---|---|---|---|---|---|---|---|---|
| EGS | 117796 | 4061 | 515 | 196 | 87 | 109 | 119 | 77 | 33 | 76 | 86 | 1 |

**Note.** Before comparing, we remove the sources with more than one HSC counterpart. The first column represents the field name, and the second ($N_{HSC}$) and third ($N_{PS1}$) columns represent the number of sources in HSC and PS1, respectively. The fourth column ($N_m$) represents the number of sources in PS1 that find more than one counterpart in the HSC within 1″, and the fifth column ($N_c$) represents the number of common sources in both catalogs. Then, out of these common sources, the next four columns represent the number of stars and galaxies in the HSC and PS1 data, respectively, where the subscript "gx" stands for galaxies and the subscript "st" stands for stars. Finally, in the last four columns we show the number of PS1 stars that are HSC galaxies (PS1$_{st}$/HSC$_{gx}$), the number of PS1 galaxies that are HSC galaxies (PS1$_{gx}$/HSC$_{gx}$), the number of PS1 stars that are HSC stars (PS1$_{st}$/HSC$_{st}$), and the number of PS1 galaxies that are HSC stars (PS1$_{gx}$/HSC$_{st}$), respectively.

(PS1$_{st}$/HSC$_{gx}$), the number of PS1 galaxies that are HSC galaxies (PS1$_{gx}$/HSC$_{gx}$), the number of PS1 stars that are HSC stars (PS1$_{st}$/HSC$_{st}$), and the number of PS1 galaxies that are HSC stars (PS1$_{gx}$/HSC$_{st}$), respectively.

In the left panel of Figure 15, we present the spatial distribution of sources familiar to HSC and PS1 across the EGS field. The middle panel depicts the ($i_{PSF} - i_{Kron}$) versus $i_{PSF}$ plot, where the significance of the orange horizontal line at 0.05 indicates the threshold for distinguishing a source as either a star or a galaxy in the PS1 data set (as described in Equation (1)). In this context, it is informative to observe the orange-colored sources, which, despite being classified as stars in the HSC data set, are deemed galaxies in the PS1 data set and reside above this threshold. Similarly, the olive sources, genuinely identified as galaxies in the HSC data, are positioned below the orange line, due to their categorization as stars in the PS1 data set. The blue sources, categorized as stars in both HSC and PS1, also lie below the orange line, while the green sources, confirmed as galaxies in both PS1 and HSC data sets, occupy a location above the orange line. Lastly, in the right panel we present the metric $d_{dev}$ for the best-fitted model after applying the optimal methodology to all stellar sources in PS1. The sources that are classified as stars in the PS1 but are classified as galaxies in HSC are relatively on the brighter end. Noticeably, there is not a significant divergence in the $d_{dev}$ metric between sources designated as stars in both PS1 and HSC and sources identified as stars in PS1 but galaxies in HSC. In addition, we note that the number of sources in the sixth column represents stars in PS1, but galaxies in HSC are likely to be quasi-stellar. Due to the small sample size, it is challenging to make conclusive remarks about the efficiency of the star–galaxy classification at this stage.

### 5.5. Fields Close to the Galactic Plane

Fields TF7, TF8, TF9, and TF10 lie close to the galactic plane. These fields also contain high amounts of dust, contributing to high extinction and differential reddening. Therefore, the optimal method cannot compute the NIR magnitudes up to the required levels of precision; see, e.g., Figure 16, which shows the comparison of the difference in the observed and computed NIR magnitudes versus the observed NIR magnitude for these test fields. Not only is the number of sources lying within a difference of 0.2 mag between the observed and computed NIR magnitudes significantly less, but there is also a shift in the scatter as a whole toward the brighter side, indicating that the computed NIR magnitudes are brighter than the observed ones. We therefore recommend using the stellar sources from the readily available NIR surveys (e.g., Visible and Infrared Survey Telescope (VISTA); Sutherland et al. 2015) in the galactic plane regions as guide stars. VISTA has a primary mirror of 4.1 m and observes the sky in NIR bands. The latest VISTA survey comprises six public surveys, e.g., UltraVISTA, VISTA Kilo-Degree Infrared Galaxy Survey (VIKING), VISTA Magellanic Survey (VMC), VISTA Variables in the Via Lactea (VVV), VISTA Hemisphere Survey (VHS), and VISTA Deep Extragalactic Observations Survey (VIDEO). While these surveys have a larger photometric depth, the data from the VVV survey (Saito et al. 2010) can be particularly helpful in obtaining information about the NIR stellar sources near the galactic plane. A study on the usability of these data can be performed in the future.

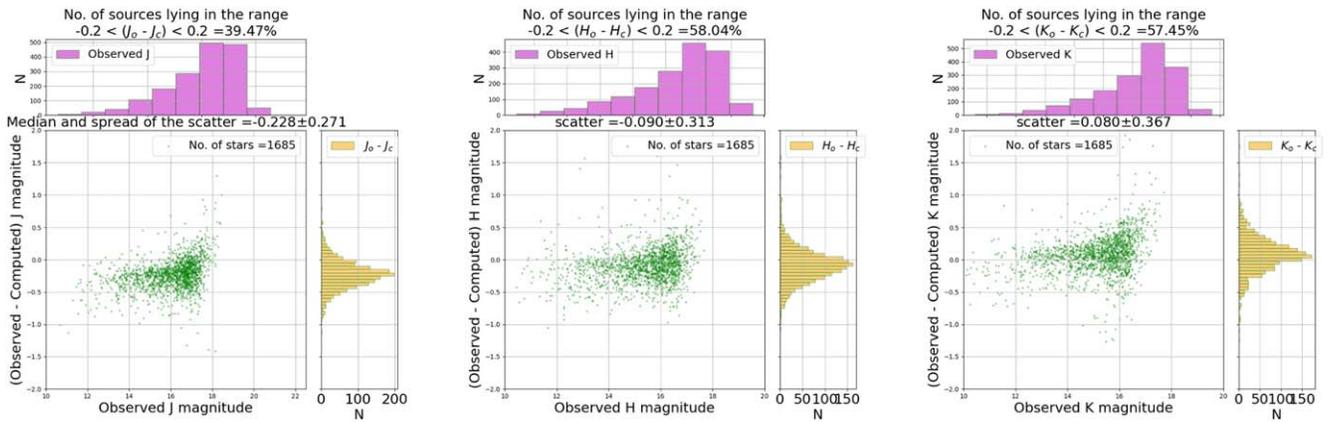

**Figure 16.** The panels compare the computed and observed NIR magnitudes when the optimal method is applied on the stack photometry for the probable stellar sources in the TF7 field, which lies close to the galactic plane.

### 5.6. Fields with Decl. $< -30°$

While the entire observable sky for TMT will be north of $-45°$ in decl., the methodology to generate IRGSC from the Pan-STARRS 3pi survey DR2 is applicable for fields located above $-30°$ decl. This is because of the coverage of the Pan-STARRS survey, which is north of $-30°$ in decl. Hence, we need to find an alternative way to generate IRGSC between $-30°$ and $-45°$ of the sky. The VHS (McMahon 2012) covers the entire sky in the Southern Hemisphere, and the stellar sources from this survey data can be used for IRGSC. Alternatively, the optical data from Vera C. Rubin Observatory (formerly known as Large Synoptic Survey Telescope (LSST)), which will observe the entire sky that is visible from Chile (Ivezić et al. 2019), could be used to generate the IRGSC. However, the optimal methodology would have to be tested and modified before computing the NIR magnitudes of the sources in the optical data of LSST.

### 5.7. Creation of the IRGSC for Additional Fields

In addition to the test fields used in this work to test and validate our methodology, we have created the IRGSCs for an additional 10 fields that belong to the Medium Deep Survey (MDS) of Pan-STARRS. MDS fields have other interesting scientific observations toward them. The generated IRGSCs for these fields are created using the 3pi survey data, as the MDS data, although deeper, are not publicly available (Chambers et al. 2016). However, whenever the data are made public, our optimal methodology can be applied, and the generated IRGSC fields can be checked for completeness and accuracy of the computed NIR magnitudes. These catalogs can be compared with those generated using the 3pi survey data. These generated IRGSCs using the 3pi survey data are available for the public on TMT GitHub repository.[18,19] It is to be noted that the MD01 field from the Pan-STARRS MDS coincides with the Andromeda galaxy. There is no IRGSC for this field because there are no 3pi survey data available.

### 6. *irgsctool*: A Python Package to Generate IRGSC

The code to generate IRGSC is publicly available in a Python package called *irgsctool*. This Python package has to be installed in a fresh environment in the system. To generate the IRGSC, the *irgsc*() class requires R.A. and decl. of the field, and then it obtains the PS1 DR2 data automatically using the *pyvo* Python module. Due to the limitation of this module, the PS1 data can be obtained for $0°.25$ only. *irgsctool* obtains the data from Gaia DR3 for the given set of coordinates. For the fields in which observed UKIDSS data are readily available, *irgsctool* can also provide a list of sources that have observed NIR magnitudes readily available. These sources can be used as NGSs. But as they are not available up to $J=22$ mag (except for some fields in the UKIDSS DXS and UDS, which are available only for a small region of the entire sky), the computed NIR magnitudes of the optical PS1 sources can be used.

### 7. Summary

In this study, we have outlined a method for constructing a partial IRGSC, building on the framework established by Subramanian et al. (2016). By synthesizing photometry from interpolated Kurucz and PHOENIX models, convolved with the PS1 and UKIDSS response functions, we have enriched our understanding of stellar populations in 20 test fields spread across the celestial sphere. These fields, situated at diverse galactic latitudes, have been selected owing to their ready accessibility for optical and NIR observations.

Through a series of methodological refinements and enhancements, we observed notable advancements in NIR magnitude computation. Specifically, the inclusion of two additional bands, $z$ and $y$, improved the accuracy of our calculations. Furthermore, we recognized that the presence of cooler giants or dwarfs, such as brown dwarfs and ultracool red dwarfs, in the photometric data posed unique challenges. Kurucz models were ineffective in computing NIR magnitudes for these sources, leading to the emergence of multiple sequences when comparing observed and computed NIR magnitudes. To address this, we introduced selective PHOENIX models to complement the interpolated Kurucz models.

---
[18] https://github.com/sshah1502/irgsc/tree/main/generated_irgsc
[19] The IRGSCs for these fields are 15′ in radius, due to the limitation of the Python package *pyvo* to retrieve the Pan-STARRS data.

By combining the interpolated Kurucz models with $T_{\rm eff} > 4000$ K and two sets of interpolated PHOENIX models within the ranges [2800 K < $T_{\rm eff}$ < 5000 K, $\log(g) > 3.0$, [Fe/H] < −1.50] and [2800 K < $T_{\rm eff}$ < 4000 K, $\log(g) < 3.0$, [Fe/H] > −0.50], we achieved robust results. The former set of PHOENIX models accurately computed NIR magnitudes for metal-poor and cooler dwarfs, while the latter set excelled with metal-rich and cool giants. However, we caution against relying on the NIR magnitudes of sources categorized as ultra−metal-poor ([Fe/H] < −2.0) and ultracool ($T_{\rm eff}$ < 3500 K), due to model limitations. We applied our optimized methodology in some fields to achieve the requisite source density for the NFIRAOS FOV. However, the source counts after star–galaxy classification fell below the required criteria in certain fields. We attribute this to our strict condition that sources must be detected in all five PS1 bands and have SNR > 5, excluding some fainter sources. While relaxing this criterion increased source count and completeness, it affected NIR magnitude accuracy and star–galaxy classification efficiency. Therefore, sources that qualify for inclusion in the IRGSC should be used with caution despite not being detected in all five bands.

Our method successfully computes NIR magnitudes up to $J = 22$ mag for all test fields, with the potential to reach deeper for specific sources. The 90% completeness of the PS1 optical input catalog typically lies in the 20–22 mag range in the $i$ band, thus setting the 90% completeness of the IRGSC in the 19–20 mag range in $J$. The limitation in the $i$ band is due to our requirement that sources must be detected in all five PS1 bands. This may change with future PS1 DR3 or PS2 releases.

While our methodology effectively computes NIR magnitudes for various sources, it does have limitations in accurately handling cooler sources, particularly in the $H$ and $K$ bands. Future improvements in SAMs may address this limitation.

In our approximation, we assume constant reddening along the line of sight for all stars within a field, which is valid for fields with low extinction. For fields away from the galactic plane (−10° < $b$ < 10°), this assumption is reasonable and helps simplify the generation of the IRGSC.

Although the star–galaxy classification criteria have been applied efficiently to the PS1 data set using HST data for multiple fields, the limited number of common sources hinders definitive conclusions. Further data from HST and missions like the James Webb Space Telescope will provide more insights.

We have provided a Python tool for generating an IRGSC, available on the TMT GitHub repository. The catalog, stored in a comma-separated values format, includes relevant astrometry information for sources matching Gaia DR3 data and the "renormalized unit weight error (RUWE)" parameter to identify single- or multiple-star systems. Additionally, the code generates essential plots, such as star–galaxy separation, error pattern comparisons, and computed versus observed NIR magnitudes.

## 8. Future Development

For future development, to improve the completeness and number density of the sources in the catalog, we suggest using more sources in the input catalog that have detection in all five optical Pan-STARRS filters to satisfy our initial selection criteria. An SNR ⩾ 5 for these sources will help better estimate the NIR magnitudes of the cooler and fainter sources. This issue will be resolved with the data release version 3.0 of Pan-STARRS, which is expected to provide more objects and better photometry of the sources. We rely on the Gaia survey for the proper-motion and parallax values of the sources in our partial IRGSC, as the current Gaia DR3 does not provide these values for all the sources, especially the fainter ones. The future releases of Gaia data are expected to have more sources and thus can be used to incorporate the proper motion and parallaxes of the fainter sources. In Section 5.3, we have already shown the changes in the results when only $g$, $r$, and $i$ bands are used to generate the IRGSC. We suggest using the sources detected in three or four optical bands and flagging them separately. These sources should be used cautiously at the time of observation. Similarly, the number of sources in the EGS is too small to draw any conclusions about the efficiency of the star–galaxy classification. We suggest more work in the future to study objects classified as stars using the PS1 data set but classified as nonstellar in the HSC data set.

In the comparison plots of the observed and computed NIR magnitudes, the large width of the scatter in the $H$ and $K$ bands could be due to the limitation of the PHOENIX models. Since the accuracy of the Kurucz models is better above $T_{\rm eff} > 4000$ K, we have to rely on the PHOENIX models to increase the source density by modeling the cooler and fainter sources. A solution to this problem can be an improvement in the existing SAM templates in the future. In addition, as already mentioned in Section 5.5, the readily available observed NIR data from the existing NIR surveys, like VISTA, which covers the whole Southern Hemisphere, can be used for the IRGSC.

Since the TMT will observe the sky up to a decl. of −45° and the Pan-STARRS survey does not cover the sky below −30° decl., the data from the Rubin Observatory can be used to compute the NIR magnitudes for the optical stellar sources in the narrow strip in between −30° and −45° decl. The optimal methodology developed here must be tested and modified to meet the requirements. Finally, an all-sky IRGSC can be generated using the Pan-STARRS DR2 data from the 3pi survey. This catalog will be based on the optimal methodology developed in this phase. This catalog can also be tested on the existing AO systems implemented on telescopes, e.g., the Keck telescope (Wizinowich et al. 2006).

## Acknowledgments

S. Subramanian acknowledges support from the Science and Engineering Research Board of India through Ramanujan Fellowship. This work made use of Astropy:[20] a community-developed core Python package and an ecosystem of tools and resources for astronomy (Astropy Collaboration et al. 2013, 2018, 2022). This work has used the NASA/IPAC Infrared Science Archive, funded by the National Aeronautics and Space Administration and operated by the California Institute of Technology, to obtain the reddening and NIR extinction coefficient values. The irgsctool package also relies on this service to obtain those values. This work also used data from the European Space Agency (ESA) mission Gaia (https://www.cosmos.esa.int/gaia), processed by the Gaia Data Processing and Analysis Consortium (DPAC, https://www.cosmos.esa.int/web/gaia/dpac/consortium). Funding for the DPAC has been provided by national institutions, in particular the institutions participating in the Gaia Multilateral Agreement. The Hubble Source Catalog (HSC) was obtained based on data from the Hubble Legacy Archive (HLA) based on observations made with the NASA/ESA Hubble Space

---

[20] http://www.astropy.org

Telescope, which is a collaboration between the Space Telescope Science Institute (STScI/NASA), the Space Telescope European Coordinating Facility (ST-ECF/ESAC/ESA), and the Canadian Astronomy Data Centre (CADC/NRC/CSA). This project acknowledges the use of *Astrolib PySynphot*, which is an object-oriented replacement for the *STSDAS SYNPHOT* synthetic photometry package in IRAF. Although this package was developed for HST, it can be utilized with other observatories. *pysynphot* simulates the photometric data and spectra as observed with the Hubble Space Telescope (HST). However, users can also incorporate their filters, spectra, and data. We thank David Andersen, TMT Project office, for constructive comments on an initial version of the report.

## Appendix A
## Schema of the Generated IRGSC

The IRGSC generated using the optimal method applied on the stack photometric data of the Pan-STARRS has various information about the sources shown in Table A1. This table describes the columns in the IRGSC generated for a particular test field. The details of the flags, e.g., infoflags, filterflags, and qualityflags, can be found in Flewelling et al. (2020). These flags indicate various values assigned to the source by Pan-STARRS, which gives further information about the nature of the source and the quality of its detection, which can help us understand more about a particular object of interest. It is to be noted that we use the stack photometric information in our analysis. Still, we use the R.A. and decl. of the source from the mean photometric database, as they are well calibrated using Gaia DR2 (Magnier et al. 2020b). For the generated IRGSC, we include the additional astrometric information from Gaia DR3, such as the proper motion and parallax. We also include a flag from Gaia DR3 called renormalized unit weight error (RUWE). RUWE indicates whether the source is single or a part of a multiple-star system. A value less than 1.3 indicates that the star is probably single.

**Table A1**
The Names of the Columns in the IRGSC

| Column Name | Description |
| --- | --- |
| ps1 objid | Object ID in PS1 data (float). |
| ps1 ra | R.A. in degrees of the source in PS1 single-epoch (mean) photometry data (float). |
| ps1 ra error | Uncertainty in R.A. in arcsec (float). |
| ps1 dec | Decl. in degrees of the source in the PS1 single-epoch (mean photometry) data (float). |
| ps1 dec error | Uncertainty in decl. in arcsec (float). |
| ps1 gpsf | PSF magnitude of the source in $g$ band (float). |
| ps1 gpsf error | Uncertainty in the PSF magnitude in $g$ band (float). |
| ps1 rpsf | PSF magnitude of the source in $r$ band (float). |
| ps1 rpsf error | Uncertainty in the PSF magnitude in $r$ band (float). |
| ps1 ipsf | PSF magnitude of the source in $i$ band (float). |
| ps1 ipsf error | Uncertainty in the PSF magnitude in $i$ band (float). |
| ps1 zpsf | PSF magnitude of the source in $z$ band (float). |
| ps1 zpsf error | Uncertainty in the PSF magnitude in $z$ band (float). |
| ps1 ypsf | PSF magnitude of the source in $y$ band (float). |
| ps1 ypsf error | Uncertainty in the PSF magnitude in $y$ band (float). |
| teff | $T_{\mathrm{eff}}$ of the best-fitted model (float). |
| logg | $\log(g)$ of the best-fitted model (float). |
| feh | [Fe/H] of the best-fitted model (float). |
| sam $g$ | Synthetic $g$ magnitude in Pan-STARRS filter(float) (in AB system). |
| sam $r$ | Synthetic $r$ magnitude in Pan-STARRS filter(float) (in AB system). |
| sam $i$ | Synthetic $i$ magnitude in Pan-STARRS filter(float) (in AB system). |
| sam $z$ | Synthetic $z$ magnitude in Pan-STARRS filter(float) (in AB system). |

**Table A1**
(Continued)

| Column Name | Description |
| --- | --- |
| sam y | Synthetic y magnitude in Pan-STARRS filter(float) (in AB system). |
| sam j | Synthetic j magnitude in UKIDSS filter(float) (in AB system). |
| sam h | Synthetic h magnitude in UKIDSS filter(float) (in AB system). |
| sam k | Synthetic k magnitude in UKIDSS filter(float) (in AB system). |
| scale factor | The scale factor computed after matching the SAM to the observed data (float). |
| scale factor error | Uncertainty in the scale factor (float). |
| $d_{dev}$ | The minimum value of $d_{dev}$ as defined in Section 2.6 (float). |
| J | Computed J in the Vega system (float). |
| J error | Uncertainty in the computed J (float). |
| H | Computed H in the Vega system (float). |
| H error | Uncertainty in the computed H (float). |
| K | Computed K in the Vega system (float). |
| K error | Uncertainty in the computed K (float). |
| gaia source id | Source ID in Gaia DR3 catalog (float). |
| gaia parallax | Parallax of the source from Gaia DR3 (float). |
| gaia parallax error | Uncertainty in the parallax (float). |
| gaia pm | Proper motion in Gaia DR3 catalog (float). |
| gaia pm ra | Proper motion along the R.A. axis in Gaia DR3 catalog (float). |
| gaia pm ra error | Uncertainty in the proper motion along the R.A. axis (float). |
| gaia pm dec | Proper motion along the decl. axis in Gaia DR3 catalog (float). |
| gaia pm dec error | Uncertainty in the proper motion along the decl. axis (float). |
| gaia ruwe | Renormalized unit weight error (float) metric in Gaia DR3. A value less than 1.4 is likely a single star (float). |
| objinfoflag | These flag values of the source in Pan-STARRS data specify whether the object is a QSO, transient, asteroid, extended, a known solar system object, etc., in nature (float). |
| objqualityflag | These flag values denote whether an object is real or a possible false positive (float). |
| ndetections | The number of times something is detected from the individual exposures (float). |
| nstackdetections | The number of stack detections after which the stack photometric measurements are done (float). |
| ginfoflag | These flags indicate the details of the g filter stack photometry (float). |
| ginfoflag2 | These flags indicate the details of the g filter stack photometry (float). |

**Table A1**
(Continued)

| Column Name | Description |
| --- | --- |
| ginfoflag3 | These flags indicate the details of the *g* filter stack photometry (float). |
| rinfoflag | These flags indicate the details of the *r* filter stack photometry (float). |
| rinfoflag2 | These flags indicate the details of the *r* filter stack photometry (float). |
| rinfoflag3 | These flags indicate the details of the *r* filter stack photometry (float). |
| iinfoflag | These flags indicate the details of the *i* filter stack photometry (float). |
| iinfoflag2 | These flags indicate the details of the *i* filter stack photometry (float). |
| iinfoflag3 | These flags indicate the details of the *i* filter stack photometry (float). |
| zinfoflag | These flags indicate the details of the *z* filter stack photometry (float). |
| zinfoflag2 | These flags indicate the details of the *z* filter stack photometry (float). |
| zinfoflag3 | These flags indicate the details of the *z* filter stack photometry (float). |
| yinfoflag | These flags indicate the details of the *y* filter stack photometry (float). |
| yinfoflag2 | These flags indicate the details of the *y* filter stack photometry (float). |
| yinfoflag3 | These flags indicate the details of the *y* filter stack photometry (float). |
| SAM name | The name of the best-fitted SAM. |

**Note.** Here pm is the acronym for proper motion, sam stands for the name of the SAM, and ps1 stands for Pan-STARRS. The information on the Pan-STARRS flag values can be found in Flewelling et al. (2020), and further information can be obtained using a CasJobs query.

## Appendix B
## Schema of the Validated IRGSC

In addition to the generation of IRGSC, *irgsctool* also validates the computed NIR magnitudes with the observed NIR UKIDSS sources obtained for the same region of the sky. The positions of the PS1 and UKIDSS sources in the sky are cross-matched up to 1″, and the *Validated IRGSC* catalog is generated. Apart from the columns described in Table A1, the validated catalog contains additional columns given in Table B1, which are mainly dependent on the UKIDSS data. This catalog cannot be generated if no observed UKIDSS data are available for the field.

Table B1
The Name of the Additional Columns in the Validated IRGSC

| Column Name | Description |
| --- | --- |
| diff_J | Difference in the observed and computed $J$ (float). |
| diff_H | Difference in the observed and computed $H$. |
| diff_K | Difference in the observed and computed $K$ (float). |
| J_UKIDSS | $J$ mag from the UKIDSS observations (petro mag) (float). |
| err_J_UKIDSS | Uncertainty in J_UKIDSS (float). |
| H_UKIDSS | $H$ mag from the UKIDSS observations (petro mag) (float). |
| err_H_UKIDSS | Uncertainty in H_UKIDSS (float). |
| K_UKIDSS | $K$ mag from the UKIDSS observations (petro mag) (float). |
| err_K_UKIDSS | Uncertainty in H_UKIDSS (float). |

**Note.** Here the observed UKIDSS NIR magnitudes are petro magnitudes.

# Appendix C
# Supplementary Figures

Figures C1–C5 contain the plots in $H$ and $K$ bands for the results obtained while developing the optimal methodology in Section 2.6.

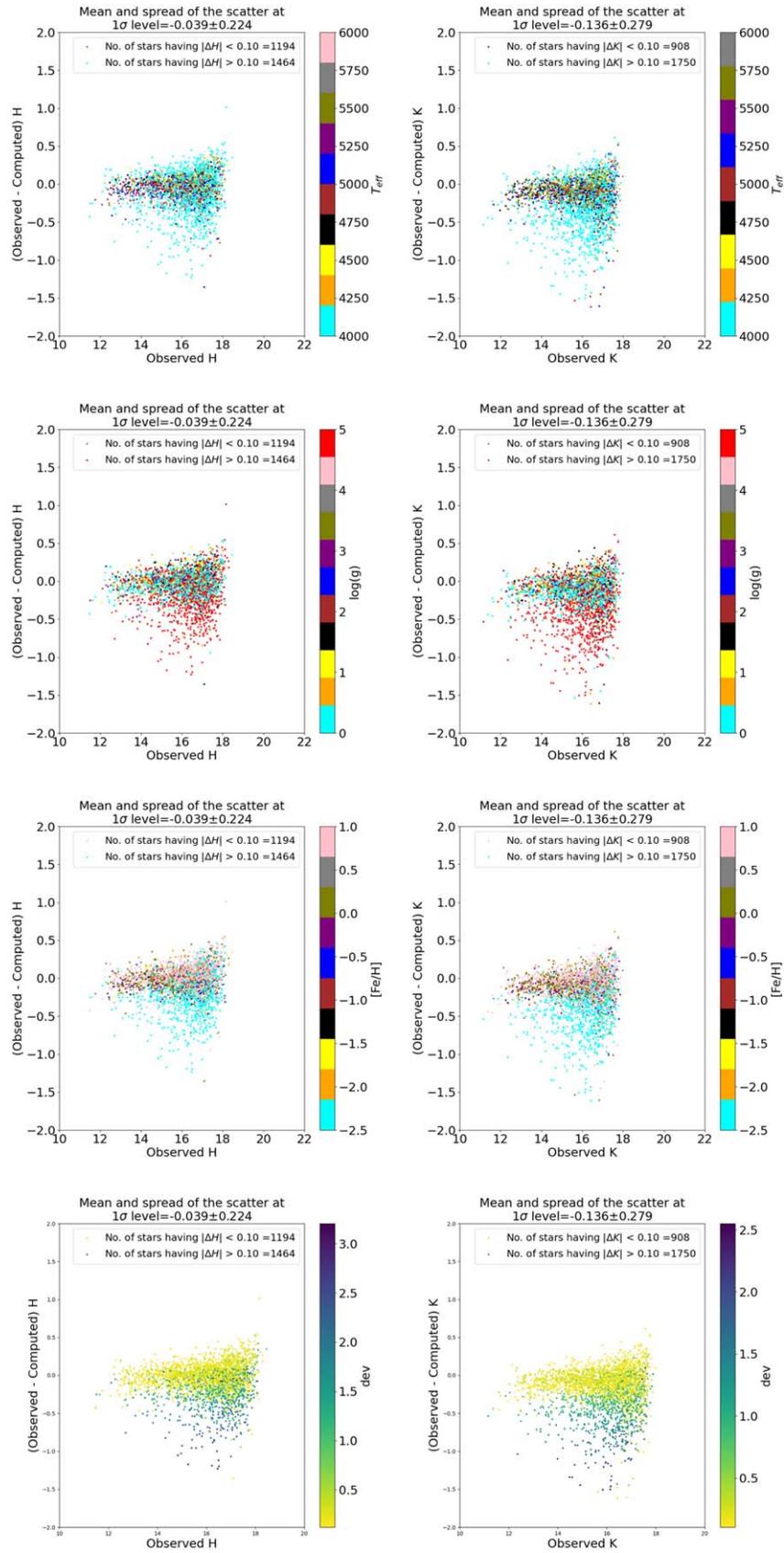

**Figure C1.** The panels show the comparison of the observed and computed *H* and *K* magnitudes when all sources are modeled with K1 models and by calculating $d_{\rm dev,min}$ for each source (see Figure 7). Each panel shows the source color-coded according to the best-fitted model parameter and dev.

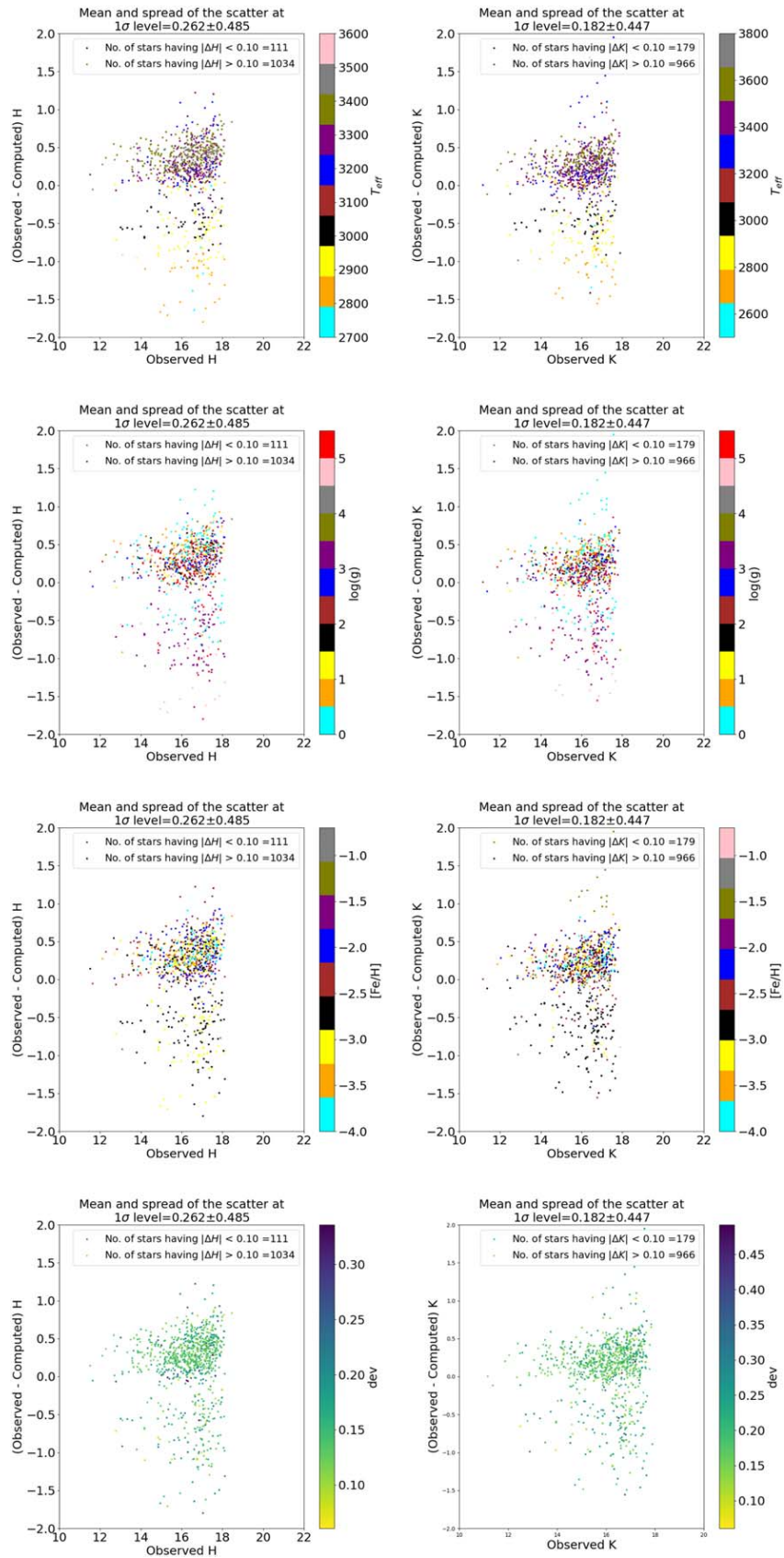

**Figure C2.** The panels show the comparison of the observed and computed *H* and *K* magnitudes when the scattered sources in the second row of Figure 7 are remodeled using P0 models.

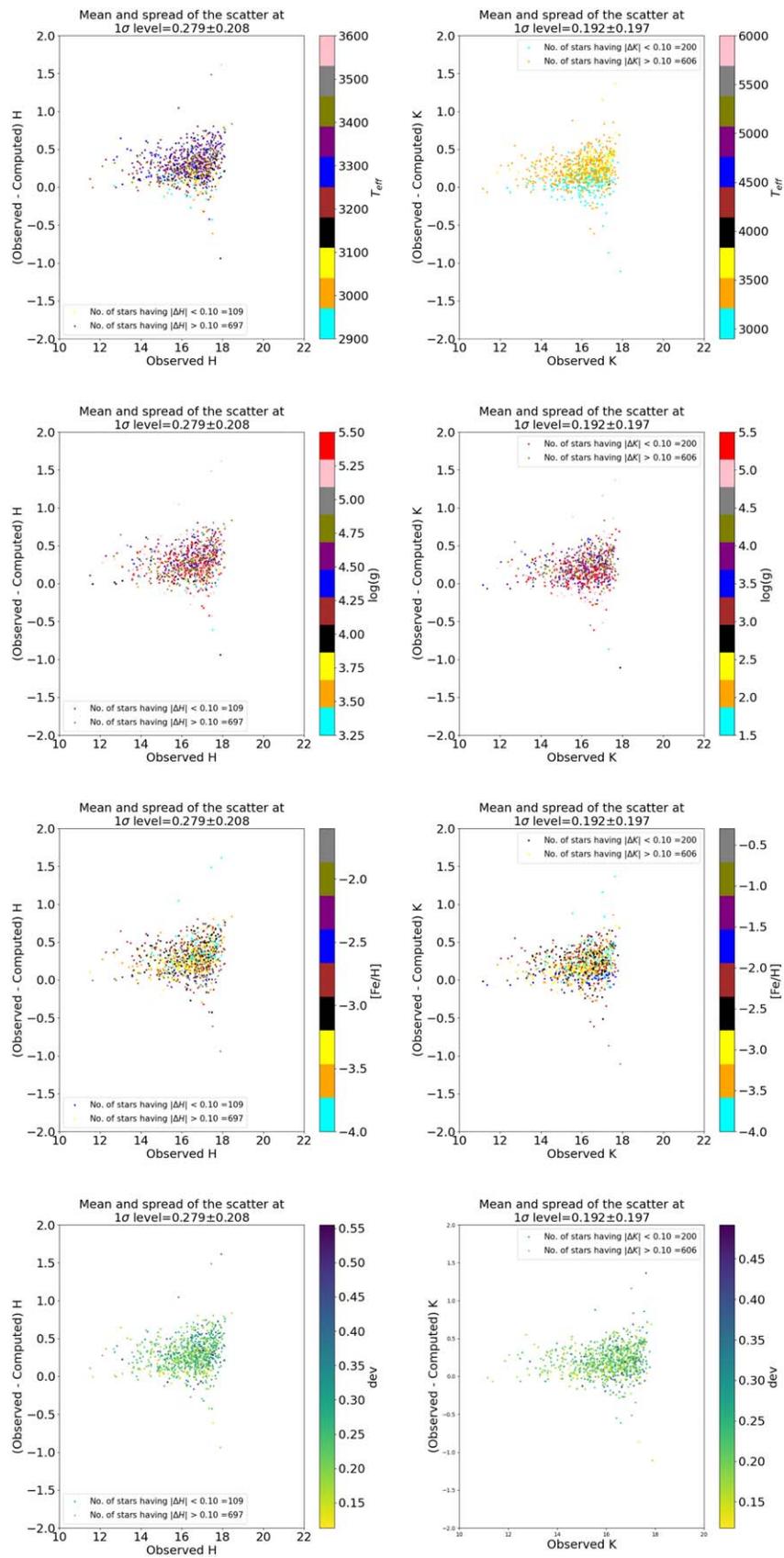

**Figure C3.** The panels show the comparison of the observed and computed $H$ and $K$ magnitudes when the scattered sources in the second row of Figure 7 are remodeled using C1 and C2 models.

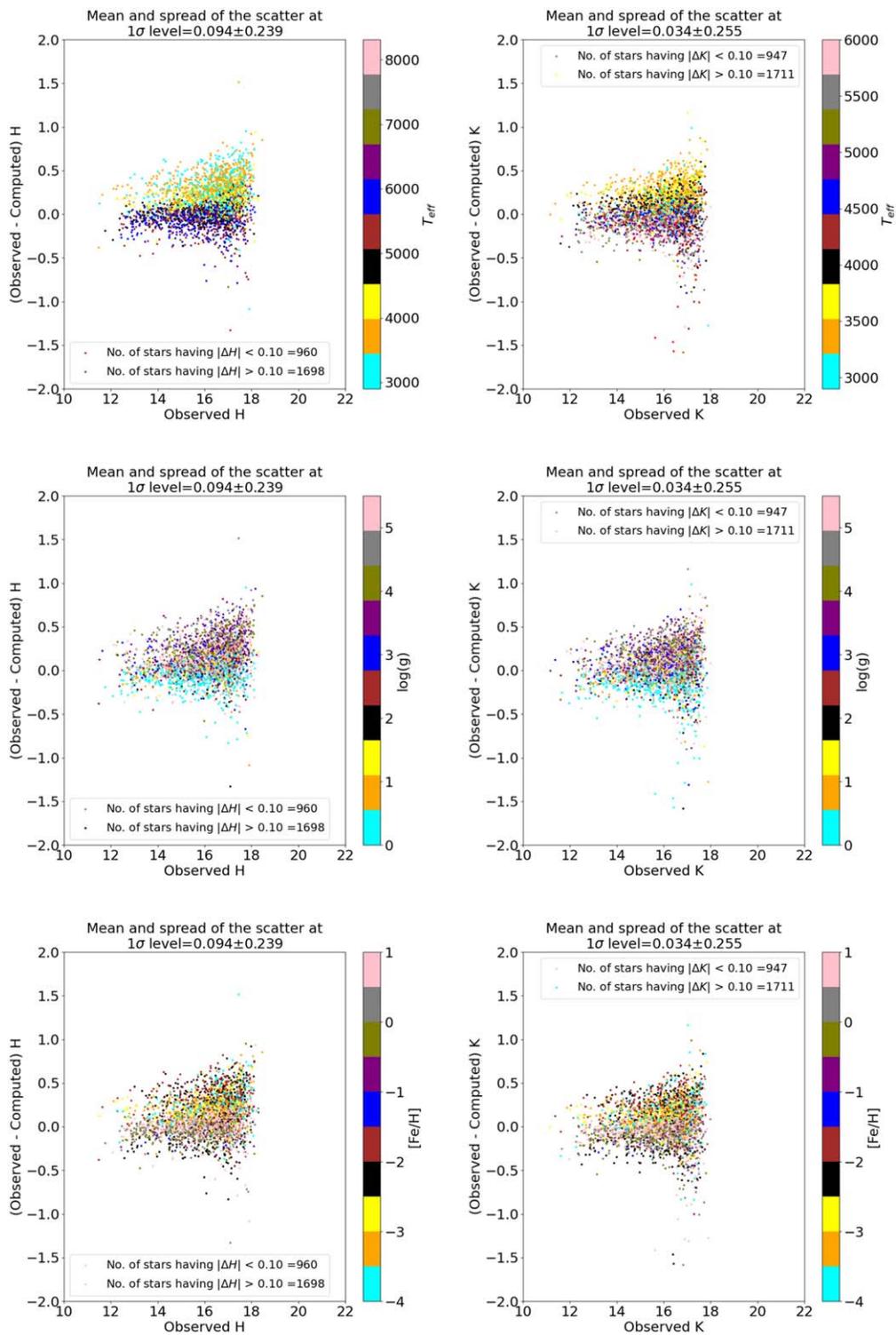

**Figure C4.** The panels show the properties of the sources in the plot, showing the comparison between observed and computed *H* and *K* magnitudes of the sources in the TF1 field when a combination of K1, C0, and C1 models is applied.

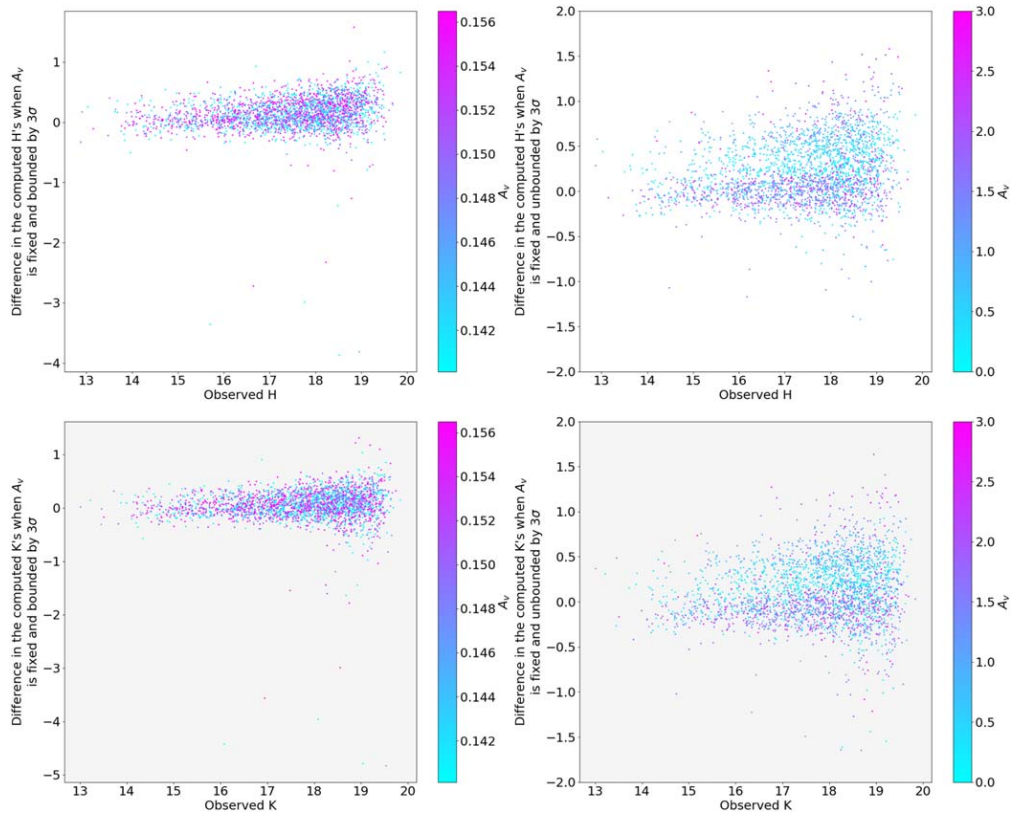

**Figure C5.** The left panels show the comparison of the difference in the computed and observed $H$ and $K$ magnitudes for the TF1 field when $A_v$ is kept free but bounded by a $3\sigma$ limit, whereas the right panels show the same but when the $A_v$ is kept free within the range of [0.0, 3.0]. The mean value of $A_v$ for the TF1 field is 0.04. The color bar shows the $A_v$ values that each source takes to minimize the $\chi_r^2$.

## ORCID iDs

Sarang Shah ⓘ https://orcid.org/0000-0003-1959-8439
G. C. Anupama ⓘ https://orcid.org/0000-0003-3533-7183
T. Sivarani ⓘ https://orcid.org/0000-0003-0891-8994
Annapurni Subramaniam ⓘ https://orcid.org/0000-0003-4612-620X

## References

Astropy Collaboration, Price-Whelan, A. M., Lim, P. L., et al. 2022, ApJ, 935, 167
Astropy Collaboration, Price-Whelan, A. M., Sipőcz, B. M., et al. 2018, AJ, 156, 123
Astropy Collaboration, Robitaille, T. P., Tollerud, E. J., et al. 2013, A&A, 558, A33
Bernstein, R. A., McCarthy, P. J., Raybould, K., et al. 2014, Proc. SPIE, 9145, 91451C
Bertone, E., Buzzoni, A., Chávez, M., & Rodríguez-Merino, L. H. 2004, AJ, 128, 829
Boyer, C., & Ellerbroek, B. 2016, Proc. SPIE, 9909, 990908
Casali, M., Adamson, A., Alves de Oliveira, C., et al. 2007, A&A, 467, 777
Castelli, F., & Kurucz, R. L. 2003, in IAU Symp. 210, Modelling of Stellar Atmospheres, ed. N. Piskunov, W. W. Weiss, & D. F. Gray (Cambridge: Cambridge Univ. Press), A20
Chambers, K. C., Magnier, E. A., Metcalfe, N., et al. 2016, arXiv:1612.05560
Crane, J. D., Herriot, G., Andersen, D. R., et al. 2018, Proc. SPIE, 10703, 107033V
Czekaj, M. A., Robin, A. C., Figueras, F., Luri, X., & Haywood, M. 2014, A&A, 564, A102
Davis, M., Guhathakurta, P., Konidaris, N. P., et al. 2007, ApJL, 660, L1
Fitzpatrick, E. L. 1999, PASP, 111, 63
Flewelling, H. A., Magnier, E. A., Chambers, K. C., et al. 2020, ApJS, 251, 7
Flicker, R. C., & Rigaut, F. J. 2002, PASP, 114, 1006
Gaia Collaboration, Vallenari, A., Brown, A. G. A., et al. 2023, AAP, 674, A1
Green, G. 2018, JOSS, 3, 695
Groth, E. J., Kristian, J. A., Lynds, R., et al. 1994, AAS Meeting Abstracts, 185, 53.09
Hauschildt, P. H., Allard, F., & Baron, E. 1999a, ApJ, 512, 377
Hauschildt, P. H., Allard, F., Ferguson, J., Baron, E., & Alexander, D. R. 1999b, ApJ, 525, 871
Hewett, P. C., Warren, S. J., Leggett, S. K., & Hodgkin, S. T. 2006, MNRAS, 367, 454
Hippler, S. 2019, JAI, 8, 1950001
Husser, T. O., Wende-von Berg, S., Dreizler, S., et al. 2013, A&A, 553, A6
Indebetouw, R., Mathis, J. S., Babler, B. L., et al. 2005, ApJ, 619, 931
Ivezić, Ž., Kahn, S. M., Tyson, J. A., et al. 2019, ApJ, 873, 111
Kron, R. G. 1980, ApJS, 43, 305
Kurucz, R. L. 1992a, RMxAA, 23, 45
Kurucz, R. L. 1992b, in IAU Symp. 149, The Stellar Populations of Galaxies, ed. B. Barbuy & A. Renzini, Vol. 149 (Dordrecht: Kluwer), 225
Kurucz, R. L. 1993, in ASP Conf. Ser. 44, IAU Coll. 138: Peculiar versus Normal Phenomena in A-type and Related Stars, ed. M. M. Dworetsky, F. Castelli, & R. Faraggiana (San Francisco, CA: ASP), 87
Larkin, J. E., Moore, A. M., Wright, S. A., et al. 2016, Proc. SPIE, 9908, 99081W
Lasker, B., Lattanzi, M. G., McLean, B. J., et al. 2008, AJ, 136, 735
Lasker, B. M., Lattanzi, M. G., McLean, B. J., et al. 2008, AJ, 136, 735
Lasker, B. M., Sturch, C. R., McLean, B. J., et al. 1990, AJ, 99, 2019
Lawrence, A., Warren, S. J., Almaini, O., et al. 2007, MNRAS, 379, 1599
Li, M., Wei, K., Tang, J., et al. 2016, Proc. SPIE, 9909, 99095Q
Magnier, E. A., Chambers, K. C., Flewelling, H. A., et al. 2020a, ApJS, 251, 3
Magnier, E. A., Sweeney, W. E., Chambers, K. C., et al. 2020b, ApJS, 251, 5
McMahon, R. 2012, in Science from the Next Generation Imaging and Spectroscopic Surveys (Garching: ESO), 37
Oke, J. B., & Gunn, J. E. 1983, ApJ, 266, 713
Ramsay, S. K., Casali, M. M., González, J. C., & Hubin, N. 2014, Proc. SPIE, 9147, 91471Z
Rhodes, J., Refregier, A., & Groth, E. J. 2000, ApJ, 536, 79
Robin, A. C., Marshall, D. J., Schultheis, M., & Reylé, C. 2012, A&A, 538, A106
Saito, R., Hempel, M., Alonso-García, J., et al. 2010, Msngr, 141, 24

Sanders, G. H. 2013, JApA, 34, 81
Schlafly, E. F., & Finkbeiner, D. P. 2011, ApJ, 737, 103
Schlegel, D. J., Finkbeiner, D. P., & Davis, M. 1998, ApJ, 500, 525
Shah, S., & Subramanian, S. 2024a, The Data for "A Partial near Infrared Guide Star Catalog for Thirty Meter Telescope Operations", v1.0.0, Zenodo, doi:10.5281/zenodo.10802894
Shah, S., & Subramanian, S. 2024b, The Software for "A Partial near Infrared Guide Star Catalog for Thirty Meter Telescope Operations", v1.0.0, Zenodo, doi:10.5281/zenodo.10797089
Skidmore, W. & TMT International Science Development Teams 2015, RAA, 15, 1945
STScI Development Team 2013, pysynphot: Synthetic Photometry Software Package, Astrophysics Source Code Library, ascl:1303.023
Subramanian, S., Subramaniam, A., Simard, L., et al. 2013, JApA, 34, 175
Subramanian, S., Subramaniam, A., Sivarani, T., et al. 2016, JApA, 37, 24
Sutherland, W., Emerson, J., Dalton, G., et al. 2015, A&A, 575, A25
Tonry, J. L., Stubbs, C. W., Lykke, K. R., et al. 2012, ApJ, 750, 99
Wang, L., Andersen, D., & Ellerbroek, B. 2012, ApOpt, 51, 3692
Waters, C. Z., Magnier, E. A., Price, P. A., et al. 2020, ApJS, 251, 4
Whitmore, B. C., Allam, S. S., Budavári, T., et al. 2016, AJ, 151, 134
Wizinowich, P. L., Le Mignant, D., Bouchez, A. H., et al. 2006, PASP, 118, 297